\documentclass[hyper]{JHEP3} % 10pt is ignored!

\usepackage{epsfig}
\usepackage{amsmath}
\usepackage{cite}
\usepackage{axodraw}
\usepackage{lscape}
\usepackage{multirow}

\newcommand{\beq}{\begin{equation}}
\newcommand{\eeq}{\end{equation}}
\newcommand{\bea}{\begin{eqnarray}}
\newcommand{\eea}{\end{eqnarray}}

\newcommand{\met}{\not \!\! E_T}
\newcommand{\mpt}{\not \!\! P_T}
\newcommand{\mptvec}{\not \!\! \vec{P}_T}

\title{$\sqrt{\hat{s}}_{min}$: a global inclusive variable for determining the mass scale of new physics in events with missing energy at hadron colliders}

\author{Partha Konar\\
        Physics Department, University of Florida,
        Gainesville, FL 32611, USA\\
        E-mail: \email{konar@phys.ufl.edu}
        }

\author{Kyoungchul Kong\\
        Theoretical Physics Department, Fermilab,
        Batavia, IL 60510, USA\\
        E-mail: \email{kckong@fnal.gov}
        }

\author{Konstantin T.~Matchev\\
        Physics Department, University of Florida,
        Gainesville, FL 32611, USA\\
        E-mail: \email{matchev@phys.ufl.edu}
        }

\received{\today}               %%
\accepted{\today}               %% These are for published papers.

\preprint{FERMILAB-PUB-08-536-T\\
          UFIFT-HEP-08-18 \\
          February 9, 2009
          } % OR: \preprint{Aaaa/Mm/Yy\\Aaa-aa/Nnnnnn}
\abstract{We propose a new global and fully inclusive variable 
$\hat{s}^{1/2}_{min}$ for determining the mass scale of new 
particles in events with missing energy at hadron colliders.
We define $\hat{s}^{1/2}_{min}$ as the minimum center-of-mass
parton level energy consistent with the measured values of the 
total calorimeter energy E and the total visible momentum $\vec{P}$.
We prove that for an arbitrary event, $\hat{s}^{1/2}_{min}$ is 
simply given by the formula $\hat{s}^{1/2}_{min}=\sqrt{E^2-P_z^2}+\sqrt{\met^2+M_{inv}^2}$,
where $M_{inv}$ is the total mass of all invisible particles 
produced in the event. We use $t\bar{t}$ production and
several supersymmetry examples
to argue that {\em the peak} in the $\hat{s}^{1/2}_{min}$
distribution is correlated with the mass threshold of the 
parent particles originally produced in the event.
This conjecture allows an estimate of the heavy superpartner 
mass scale (as a function of the LSP mass) in a completely 
general and model-independent way, and {\em without the need for
any exclusive event reconstruction}. In our SUSY examples
of several multijet plus missing energy signals, the accuracy
of the mass measurement based on $\hat{s}^{1/2}_{min}$
is typically at the percent level, and never worse than $10\%$.
After including the effects of initial state radiation 
and multiple parton interactions, the precision gets worse, 
but for heavy SUSY mass spectra remains $\sim 10\%$.
}


\begin{document} 

%%%%%%%%Section #1  %%%%%%%%%%%%%%%%%%%%%%%%%%%%%%%%%%%
%%%%%%%%%%%%%%%%%%%%%%%%%%%%%%%%%%%%%%%%%%%%%%%%%%%%%%%
\section{Introduction}
\label{sec:intro}

The ongoing Run II of the Fermilab Tevatron and the
imminent run of the Large Hadron Collider (LHC)
at CERN are on the hunt for new physics beyond the
Standard Model (BSM) at the TeV scale. Arguably the
most compelling {\em phenomenological} evidence for
BSM particles and interactions at the TeV scale is
provided by the dark matter problem \cite{Bertone:2004pz},
whose solution requires new particles and interactions BSM.
A typical particle dark matter candidate does not interact
in the detector and can only manifest itself as missing energy.
At hadron colliders, where the total center of mass energy 
in each event is unknown, the missing energy is inferred from 
the imbalance of the total transverse momentum of the detected
visible particles, and is commonly referred to as 
``missing transverse energy'' (MET). 
The dark matter problem therefore greatly motivates the
study of MET signatures at the Tevatron and
the LHC \cite{Hubisz:2008gg}. 

While the MET class of BSM signatures is probably the best motivated one
from a theoretical point of view, it is also among the most challenging
from an experimental point of view. On the one hand, to get a good 
MET measurement, one needs to have all detector components 
working properly, since the mismeasurement of any one single type
of objects would introduce fake MET. In addition, there are complications 
from cosmics, pile-up, beam halo, noise, etc. Therefore, establishing a 
MET signal due to some new physics is a highly non-trivial task
\cite{Hubisz:2008gg,Ball:2007zza}.

At the same time, {\em interpreting} a missing energy signal 
of new physics is quite challenging as well. The main stumbling block 
is the fact that we are missing some of the kinematical information 
from each event, namely the energies and momenta of the missing 
invisible particles. What is worse, a priori we cannot be certain
about the exact number of missing particles in the event, or 
their identity, e.g. are they SM neutrinos, new BSM dark matter particles,
or some combination of both? These difficulties are illustrated in 
Fig.~\ref{fig:metevent}, where we show the generic topology 
of the missing energy events that we are considering in this paper.
%%%%%%%%%%%%%%%%%%% BEGIN FIGURE %%%%%%%%%%%%%%%%%%%
\FIGURE[ht]{
%\begin{center}
{
\unitlength=1.5 pt
\SetScale{1.5}
\SetWidth{1.0}      % line    size control
\normalsize    %  letter  size control
{} \qquad\allowbreak
%  diagram # 1
\begin{picture}(250,200)(0,0)
\SetColor{Gray}
% visible 
\Line( 13,185)(130,185)
\Line( 50,150)(130,150)
\Line( 50,140)(130,140)
\Line( 50,130)(130,130)
\Line( 50,120)(130,120)
\Text(135,120)[l]{\Black{$X_1$}}
\Text(135,130)[l]{\Black{$X_2$}}
\Text(135,140)[l]{\Black{$X_3$}}
\Text(135,150)[l]{\Black{$X_4$}}
\Text(135,185)[l]{\Black{$X_{n_{vis}}$}}
% neutrinos
\DashLine(50,110)(130,110){2}
\DashLine(50, 85)(130, 85){2}
\DashLine(50, 75)(130, 75){2}
\Text(135,110)[l]{$\chi_{n_{inv}}$}
\Text(135, 85)[l]{$\chi_{n_{\chi}+2}$}
\Text(135, 75)[l]{$\chi_{n_{\chi}+1}$}
% initial state
\Line( 10,190)(55,120)
\Line( 10, 10)(55, 80)
\Text(20,155)[c]{\Black{$p(\bar{p})$}}
\Text(20, 45)[c]{\Black{$p(\bar{p})$}}
\DashLine(95,180)(95,155){1}
\DashLine(95,105)(95, 90){1}
\CArc(10,152.5)(150,-14,14)
\LongArrow(167,152.5)(197,152.5)
\SetColor{Red}
\DashLine(95, 60)(95, 45){1}
\CArc(15,   70)(144,-17,17)
\LongArrow(165,70)(195,70)
\SetWidth{1.2}      % line    size control
\DashLine(50,65)(130,65){2}
\DashLine(50,40)(130,40){2}
\DashLine(50,30)(130,30){2}
\Text(135,65)[l]{\Red{$\chi_{n_\chi}$}}
\Text(135,40)[l]{\Red{$\chi_2$}}
\Text(135,30)[l]{\Red{$\chi_1$}}
\CBoxc(230,152.5)(50,20){Black}{Yellow}
\Text(230,152.5)[c]{$E,P_x,P_y,P_z$}
\CBoxc(215, 70)(25,20){Red}{Yellow}
\Text(215,70)[c]{\Red{$\mptvec$}}
\COval(50,100)(75,15)(0){Blue}{Green}
\end{picture}
}
\caption{The generic event topology under consideration in this paper.
Black (red) lines correspond to SM (BSM) particles.
The solid lines denote SM particles $X_i$, $i=1,2,\ldots, n_{vis}$,
which are visible in the detector, e.g.~jets, electrons, muons and photons.
The SM particles may originate either from initial 
state radiation (ISR), or from the hard scattering and subsequent 
cascade decays (indicated with the green-shaded ellipse).
The dashed lines denote neutral stable particles 
$\chi_i$, $i=1,2,\ldots, n_{inv}$, which are invisible in the detector.
In general, the set of invisible particles consists of some number 
$n_{\chi}$ of BSM particles (indicated with the red dashed lines),
as well as some number $n_{\nu}=n_{inv}-n_{\chi}$ of SM neutrinos 
(denoted with the black dashed lines). The identities and the masses
$m_i$ of the BSM invisible particles $\chi_i$, ($i=1,2,\ldots,n_{\chi}$)
do not necessarily have to be all the same, i.e.~we allow for the
simultaneous production of several {\em different} species of dark 
matter particles. The global event variables describing the visible
particles are: the total energy $E$, the transverse components 
$P_x$ and $P_y$ and the longitudinal component $P_z$ 
of the total visible momentum $\vec{P}$. The only experimentally available
information regarding the invisible particles is the missing transverse
momentum $\mptvec$.
}
\label{fig:metevent} 
%\end{center}
}
%%%%%%%%%%%%% END OF FIGURE ################
As can be seen from the figure, we are imagining
a completely general setup -- each event will contain
a certain number $n_{vis}$ of Standard Model (SM) particles 
$X_i$, $i=1,2,\ldots, n_{vis}$, which are {\em visible} in the detector,
i.e. their energies and momenta are in principle measured.
Examples of such visible SM particles 
are the basic reconstructed objects, e.g.~jets, photons, electrons and muons.
The visible particles $X_i$ are denoted in Fig.~\ref{fig:metevent} 
with solid black lines and
may originate either from initial state radiation (ISR), 
or from the hard scattering and subsequent 
cascade decays (indicated with the green-shaded ellipse).
On the other hand, the missing energy $\met$ 
(or more appropriately, the missing transverse
momentum $\mptvec$) will arise from a certain number $n_{inv}$
of stable neutral particles $\chi_i$, $i=1,2,\ldots, n_{inv}$, 
which are {\em invisible} in the detector. 
In general, the set of invisible particles in any event 
will consist of a certain number 
$n_{\chi}$ of BSM particles (indicated with the red dashed lines),
as well as a certain number $n_{\nu}=n_{inv}-n_{\chi}$ 
of SM neutrinos (denoted with the black dashed lines).
The missing energy measurement alone does not tell us 
the number $n_{inv}$ of missing particles, nor how many of them
are neutrinos and how many are BSM (dark matter) particles.
Notice that in this general setup the identities and the masses
$m_i$ of the BSM invisible particles $\chi_i$, 
($i=1,2,\ldots,n_{\chi}$)
do not necessarily have to be all the same, i.e.~we allow for the
simultaneous production of several {\em different} species of dark 
matter particles \cite{Hur:2007ur,Cao:2007fy,SungCheon:2008ts,Hur:2008sy}. 
On the other hand, we shall always take the neutrino masses to be zero
\beq
m_i=0, \quad{\rm for}\ i=n_\chi+1,n_\chi+2,\ldots,n_{inv}\ .
\label{zeromnu}
\eeq

Most previous studies of MET signatures
have assumed a particular BSM scenario and investigated its 
consequences in a rather model-dependent setup. The results
from those studies would seem to indicate that in order to make
any progress towards determining what kind of new physics is 
being discovered, and in particular towards mass and spin measurements,
one must attempt at least some partial reconstruction 
of the events, by assuming a particular production mechanism, and then 
identifying the decay products from a suitable decay chain
\cite{Hinchliffe:1996iu,Lester:1999tx,
Bachacou:1999zb,Hinchliffe:1999zc,Tovey:2000wk,Allanach:2000kt,Barr:2003rg,Nojiri:2003tu,Barr:2004ze,
Goto:2004cpa,Kawagoe:2004rz,Gjelsten:2004ki,Gjelsten:2005aw,Birkedal:2005cm,Smillie:2005ar,Datta:2005zs,
Miller:2005zp,Barr:2005dz,Meade:2006dw,Lester:2006yw,Athanasiou:2006ef,Wang:2006hk,Gjelsten:2006tg,
Matsumoto:2006ws,Cheng:2007xv,Lester:2007fq,Cho:2007qv,Gripaios:2007is,Barr:2007hy,
Cho:2007dh,Ross:2007rm,Nojiri:2007pq,Huang:2008ae,
Nojiri:2008hy,Tovey:2008ui,Nojiri:2008ir,Cheng:2008mg,Cho:2008cu,Serna:2008zk,Bisset:2008hm,
Barr:2008ba,Kersting:2008qn,Nojiri:2008vq,Burns:2008cp,Cho:2008tj,Cheng:2008hk,Burns:2008va,
Barr:2008hv,Graesser:2008qi}.
In doing so, one inevitably encounters a combinatorial problem
whose severity depends on the new physics model and the type of 
discovery signature. For example, complex event topologies with 
a large number $n_{vis}$ of visible particles, 
and/or a large number of jets but few or no leptons,  
will be rather difficult to decipher, especially in the early data.
Therefore, it is fair to ask whether one can say something about
the newly discovered physics and in particular about its mass scale,
using only inclusive and global\footnote{Here and throughout the paper, 
we use the term ``global'' from an experimentalist's point of view.
Strictly speaking, the detectors are not fully hermetic, hence no
variable can be truly global in the theorist's sense.} 
event variables, before attempting any event reconstruction. 

In this paper, therefore, we shall concentrate on the most general topology
exhibited in Fig.~\ref{fig:metevent} and we shall make no further 
assumptions about the underlying event structure. 
For example, we shall not specify anything about the production mechanism.
In particular, we shall {\em not} make the usual assumption 
that the BSM particles are pair produced and, consequently, 
that there are two and only two BSM decay chains resulting in
$n_\chi=2$ identical dark matter particles with equal masses
$m_1=m_2$. Accordingly, we shall not make any attempt to group
the observed SM objects $X_i$, $i=1,2,\ldots, n_{vis}$, into 
subsets corresponding to individual decay chains. Furthermore,
we shall in principle 
allow for the presence of SM neutrinos which could contribute 
towards the measured MET. In this sense our approach 
will be completely general and model-independent.

Given this very general setup, our first goal will be to
define a global event variable which is sensitive to the 
mass scale of the particles that were originally 
produced in the event of Fig.~\ref{fig:metevent},
or more generally, to the typical energy scale of the event.
Since we are not attempting any event reconstruction, this
variable should be defined only in terms of the global event
variables describing the visible particles $X_i$, namely, 
the {\em total} energy $E$ in the event, the transverse components 
$P_x$ and $P_y$ and the longitudinal component $P_z$ 
of the {\em total} visible momentum $\vec{P}$ in the event. 
In the same spirit, the only experimentally available information 
regarding the invisible particles that we are allowed to use 
is the missing transverse momentum $\mptvec$ (see Fig.~\ref{fig:metevent}).
Of course, the missing transverse momentum $\mptvec$ is related to the
transverse components $P_x$ and $P_y$ of the total visible 
momentum $\vec{P}$ as
\beq
\mptvec = - \left( P_x \vec{e}_x + P_y \vec{e}_y \right) = - \vec{P}_T,
\eeq
so that we can use $\mptvec$ and $\vec{P}_T \equiv P_x \vec{e}_x + P_y \vec{e}_y$ 
interchangingly. Then, the commonly used missing energy $\met$ is nothing but
the magnitude $\mpt$ of the measured missing momentum $\mptvec$:
\beq
\met \equiv\ \mpt = P_T = \sqrt{ P_x^2 + P_y^2} \ .
\label{met}
\eeq 

The main idea of this paper is to propose a new global and inclusive variable
$\hat{s}_{min}$ defined as follows. $\hat{s}_{min}$ is simply the {\em minimum}
value of the parton-level Mandelstam variable $\hat{s}$ 
which is consistent with the observed set of $E$, $P_z$ and $\mpt$ 
in a given event\footnote{In what follows, instead of $\mpt$ we 
choose to use the more ubiquitous $\met$, since the two are 
essentially the same, see (\ref{met}).}. Correspondingly, its square root
$\hat{s}^{1/2}_{min}$ is the {\em minimum} parton level 
center-of-mass energy, which is required in order to explain
the observed values of $E$, $P_z$ and $\met$.
Our main result, derived below in Section~\ref{sec:derivation}, 
is the relation expressing the so defined $\hat{s}^{1/2}_{min}$ in terms 
of the measured global and inclusive quantities $E$, $P_z$ and $\met$.
In Section~\ref{sec:derivation} we shall prove that
$\hat{s}^{1/2}_{min}$ is always given by the formula
\beq
\hat{s}^{1/2}_{min}(M_{inv}) \equiv \sqrt{E^2-P_z^2}+\sqrt{\met^2+M_{inv}^2}\ ,
\label{smin_def}
\eeq
where the mass parameter $M_{inv}$ is nothing but 
the total mass of all invisible particles in the event:
\beq
M_{inv} \equiv \sum_{i=1}^{n_{inv}} m_i = \sum_{i=1}^{n_{\chi}} m_i\ ,
\label{minv}
\eeq
and the second equality follows from the assumption 
of vanishing neutrino masses (\ref{zeromnu}).

As can be seen from its defining equation (\ref{smin_def}),
the variable $\hat{s}^{1/2}_{min}$ is actually a function 
of the unknown mass parameter $M_{inv}$. This is the price 
that we will have to pay for the model-independence of our setup.
This situation is very similar to the case of the Cambridge 
$M_{T2}$ variable \cite{Lester:1999tx,Barr:2003rg,Cho:2007qv,Gripaios:2007is,%
Barr:2007hy,Cho:2007dh,Cho:2008cu,Cho:2008tj,Cheng:2008hk} and its various 
cousins \cite{Lester:2007fq,Ross:2007rm,Nojiri:2007pq,Nojiri:2008hy,Tovey:2008ui,%
Serna:2008zk,Barr:2008ba,Nojiri:2008vq,Burns:2008va,Barr:2008hv}, 
which are also defined in terms of the unknown test mass 
of a missing BSM particle. However, the Cambridge $M_{T2}$ variable
is a much more model-dependent quantity, since it requires
the identification of two separate decay chains in the events.
Furthermore, in some special cases (more precisely, those of
$M_{T2}^{(n,n,n-1)}$ in the language of \cite{Burns:2008va})
$M_{T2}$ is essentially a purely transverse quantity, and in this 
sense would not make full use of all of the available information 
in the event. In contrast, our variable $\hat{s}^{1/2}_{min}$ is defined
in a fully inclusive manner, and uses the longitudinal
event information as well. 

After deriving our main result (\ref{smin_def}) in Section~\ref{sec:derivation},
we devote the rest of the paper to studies of its properties.
For example, in Section~\ref{sec:compare} we shall compare 
$\hat{s}^{1/2}_{min}$ to some other global and inclusive variables 
which have been considered as measures of the mass scale of the 
new particles: $H_T$ \cite{Tovey:2000wk},
the total visible invariant mass $M$ \cite{Hubisz:2008gg}, 
the missing transverse energy $\met$,
the total energy $E$, and the total transverse energy
$E_T$ in the event. We shall use several examples from SM $t\bar{t}$ production, as
well as supersymmetry (SUSY), to demonstrate that among all those possibilities, 
the variable $\hat{s}^{1/2}_{min}$ is the one which is best correlated with
the mass scale of the produced particles, even when we conservatively 
set the unknown mass parameter $M_{inv}$ to zero.
In Section~\ref{sec:mass} we shall investigate the 
dependence of the $\hat{s}^{1/2}_{min}$ variable on the a priori unknown 
mass parameter $M_{inv}$, using
conventional SUSY pair-production for illustration.
We shall find a very interesting result: when the 
parameter $M_{inv}$ happens to be equal to its true value, 
the {\em peak} in the $\hat{s}^{1/2}_{min}$ distribution 
is surprisingly close to the SUSY mass threshold.
This correlation persists even when the
two SUSY particles produced in the hard scattering are very 
different, for example, in associated gluino-LSP production. 
This observation opens up the possibility of a new,
all inclusive and completely model-independent 
measurement of the mass scale of the new (parent) particles
produced in the event: we simply read off the location
of the peak in the $\hat{s}^{1/2}_{min}$ distribution, and
interpret it as the mass threshold of the parent particles.
Because of the intrinsic dependence on the unknown mass parameter
$M_{inv}$, the method only provides a relation between the mass 
of the parent particle and the mass of the dark matter particle, 
just like the method of the Cambridge $M_{T2}$ variable \cite{Lester:1999tx}.
However, unlike the $M_{T2}$ endpoint measurements, our measurement is 
based on an all-inclusive global variable, and does not require 
any event reconstruction at all. It is worth noting that
since we are correlating a physics parameter to the {\em peak}, 
rather than the {\em endpoint} of an observed distribution, our
measurement will be less prone to errors due to finite statistics, 
detector resolution, finite width effects etc., which
represents another important advantage of the $\hat{s}^{1/2}_{min}$ variable. 
The accuracy of our new mass measurement method is investigated 
quantitatively in Sections~\ref{sec:thr} and \ref{sec:ISR}. 
Our discussion in Sections~\ref{sec:compare}, \ref{sec:mass}
and \ref{sec:thr}, while demonstrating the usefullness of the 
$\hat{s}^{1/2}_{min}$ variable, will be limited to an ideal case, 
where the effects from initial state radiation (ISR), multiple parton 
interactions (MPI) and pile-up are negligible.
In Section~\ref{sec:ISR} we investigate the adverse effect
of those latter factors on the $\hat{s}^{1/2}_{min}$ measurement
in a realistic experimental environment
and discuss different approaches for minimizing their impact.
In Section~\ref{sec:summary} we summarize our main points and conclude.

\section{Derivation of $\hat{s}^{1/2}_{min}$}
\label{sec:derivation}

In this section we shall derive the general formula (\ref{smin_def})
advertised in the Introduction. Before we begin, let us introduce 
some notation. We shall denote the three-momenta 
of the invisible particles $\chi_i$, $i=1,2,\ldots,n_{inv}$, 
with $\vec{p}_i$, or in components $p_{ix}$, $p_{iy}$ and $p_{iz}$.
As usual, we choose the $z$-axis along the beam direction, so that
$p_{ix}$ and $p_{iy}$ are the components of the 
transverse momentum $\vec{p}_{iT}$. As already mentioned in 
the Introduction, the masses of the invisible particles will be
denoted by $m_i$.

Our starting point will be the expression for the parton-level
Mandelstam variable $\hat{s}$ for the event depicted in Fig.~\ref{fig:metevent}:
\bea
\hat{s} &=& \left(\, E+\sum_{i=1}^{n_{inv}}\sqrt{m_i^2+\vec{p}_i^{\,2}} \, \right)^2
          - \left(\, \vec{P}+\sum_{i=1}^{n_{inv}} \vec{p}_i \, \right)^2 
\nonumber  \\ [2mm]
&=&  \left(\, E+\sum_{i=1}^{n_{inv}}\sqrt{m_i^2+\vec{p}_{iT}^{\,2} + p_{iz}^2} \, \right)^2
          - \left(\, \vec{P}_T+\sum_{i=1}^{n_{inv}} \vec{p}_{iT} \, \right)^2 
          - \left(\, P_z+\sum_{i=1}^{n_{inv}} p_{iz} \, \right)^2 .
\label{shat} 
\eea
The invisible particle momenta $\vec{p}_i$ are not measured and are therefore
unknown. However, they are subject to the missing energy constraint:
\beq
\sum_{i=1}^{n_{inv}} \vec{p}_{iT} = \ \mptvec = -\vec{P}_T\ ,
\label{mptvec_constraint}
\eeq
which causes the second term in (\ref{shat}) to vanish and we arrive at
a simpler version of (\ref{shat})
\beq
\hat{s} =
\left(\, E+\sum_{i=1}^{n_{inv}}\sqrt{m_i^2+\vec{p}_{iT}^{\,2} + p_{iz}^2} \, \right)^2
          - \left(\, P_z+\sum_{i=1}^{n_{inv}} p_{iz} \, \right)^2 .
\label{shat2} 
\eeq
We see that the expression for $\hat{s}$ is a function of a total
of $3n_{inv}$ variables $\vec{p}_i$ which are subject to the 
2 constraints (\ref{mptvec_constraint}). Given that we are missing
so much information about the missing momenta $\vec{p}_i$, it is clear that
there is no hope of determining $\hat{s}$ {\em exactly} from experiment,
and the best one can do is to use some kind of an approximation for it.
For example, Ref.~\cite{Cho:2008tj} recently proposed to 
approximate the real values of the missing momenta $\vec{p}_i$ with
the values that determine the event $M_{T2}$ variable. 
However, constructing any $M_{T2}$ variable requires 
one to make certain model-dependent assumptions about the underlying 
topology of the event, and furthermore, for very complex events,
with large $n_{vis}$, the associated combinatorial problem 
will become quite severe. Therefore, here we shall use a different, 
more model-independent approach. The key is to realize that
the function $\hat{s}$ has an absolute global minimum $\hat{s}_{min}$, 
when considered as a function of the unknown variables $\vec{p}_i$.
Therefore, we choose to approximate the real values of the 
missing momenta with the values corresponding to the global minimum $\hat{s}_{min}$.
The minimization of the function (\ref{shat2}) with respect 
to the variables $\vec{p}_i$, subject to the constraint 
(\ref{mptvec_constraint}), is rather straightforward. %, albeit tedious. 
The global minimum is obtained for
\bea
\vec{p}_{iT} &=& \frac{m_i}{M_{inv}}\, \mptvec\ , \label{ptsol} \\ [2mm]
p_{iz} &=& \frac{m_i P_z}{\sqrt{E^2-P_z^2}} \sqrt{1+\frac{\mpt^2}{M_{inv}^2}}\ ,
\label{pzsol}
\eea
where the parameter
\beq
M_{inv} \equiv \sum_{i=1}^{n_{inv}} m_i = \sum_{i=1}^{n_{\chi}} m_i
\eeq
was already defined in (\ref{minv})
and represents the total mass of all invisible particles in the event.
Since the neutrinos are massless, $M_{inv}$ only counts the 
masses of the BSM invisible particles which are present in the event. 
Substituting (\ref{ptsol}) and (\ref{pzsol}) into (\ref{shat2}) and simplifying, 
we get the minimum value $\hat{s}_{min}$ of the function (\ref{shat2}) to be
\beq
\hat{s}_{min}(M_{inv}) = \left(\sqrt{E^2-P_z^2}+\sqrt{\mpt^2+M_{inv}^2}\right)^2\ .
\label{shat3}
\eeq
Since the right-hand side is a complete square, it is convenient to 
take the square root of both sides and consider instead
\beq
\hat{s}^{1/2}_{min}(M_{inv}) 
= \sqrt{E^2-P_z^2}+\sqrt{\mpt^2+M_{inv}^2} \ ,
\label{smin_def_again}
\eeq
which can be equivalently rewritten in terms of the missing energy $\met$ as
\begin{equation}
\hat{s}^{1/2}_{min}(M_{inv}) 
= \sqrt{E^2-P_z^2}+\sqrt{\met^2+M_{inv}^2}\ ,
\label{smin_def_met}
\end{equation}
completing the proof of (\ref{smin_def}). 

A few comments regarding the variable $\hat{s}^{1/2}_{min}$ defined
in (\ref{smin_def_met}) are in order. Perhaps the most striking
feature of $\hat{s}^{1/2}_{min}$ is its simplicity: the result 
(\ref{smin_def_met}) holds for completely general types of events,
with any number and/or types of missing particles. Clearly,
$\hat{s}^{1/2}_{min}$ itself is both a global and an inclusive 
variable, since it is defined in terms of the global and inclusive 
event quantities $E$, $P_z$ and $\met$, which do not require 
any explicit event reconstruction. It is easy to see that the expression
(\ref{smin_def_met}) is invariant under longitudinal boosts, 
since it depends on the quantities $E^2-P_z^2$, $\met$ and $M_{inv}$,
all three of which are invariant under such boosts. 
Also notice that $\hat{s}^{1/2}_{min}$ has units of energy and thus provides some 
measure of the energy scale in the event, and can be directly 
compared to other popular energy-scale variables 
(see Section~\ref{sec:compare} below). In the remainder of 
this paper we shall investigate in more detail the properties 
of the new variable (\ref{smin_def_met}). 

\section{Comparison between $\hat{s}^{1/2}_{min}$ and other global inclusive variables}
\label{sec:compare}

The immediate question after the discovery of a MET signal of new physics 
at the Tevatron or LHC, will be: ``What is the energy scale of the new physics?''.
We shall now argue that our global inclusive variable $\hat{s}^{1/2}_{min}$
from (\ref{smin_def_met}) provides a first,
relatively quick answer to this question, which will turn out to be 
surprisingly accurate, given that we are not attempting any 
event reconstruction or modelling of the new physics. 
Of course, one might do better by considering exclusive signatures
and applying the usual tricks for mass measurements, but chances are
that this will require some time. It is therefore worth investigating
how much information one can get from totally inclusive measurements 
like (\ref{smin_def_met}) which should be available from very early on. 

To set up the subsequent discussion, let us introduce the different 
global variables from Fig.~\ref{fig:metevent} which will be experimentally 
accessible. The total visible energy $E$ is simply
\beq
E = \sum_\alpha E_\alpha\ ,
\label{E}
\eeq
where we use the index $\alpha$ to label the calorimeter towers,
and $E_\alpha$ is the energy deposit in the $\alpha$ tower\footnote{We 
ignore the difference in the segmentation of the hadronic and electromagnetic 
calorimeters, and for $E_\alpha$ simply add up the HCAL and ECAL energy deposits.}.
As usual, since muons do not deposit significantly in 
the calorimeters, the measured $E_\alpha$ should first be corrected
for the energy of any muons which might be present in the event
and happen to pass through the corresponding tower $\alpha$. 
The three components of the total visible momentum $\vec{P}$ are
\bea
P_x &=& \sum_\alpha E_\alpha \sin\theta_\alpha \cos\varphi_\alpha\ ,  \label{Px}\\ [2mm]
P_y &=& \sum_\alpha E_\alpha \sin\theta_\alpha \sin\varphi_\alpha\ ,  \label{Py} \\ [2mm]
P_z &=& \sum_\alpha E_\alpha \cos\theta_\alpha\ ,  \label{Pz}
\eea
where $\theta_\alpha$ and $\varphi_\alpha$ are correspondingly 
the azimuthal and polar angular coordinates of the $\alpha$ calorimeter tower. 
The total transverse energy $E_T$ is
\beq
E_T \equiv \sum_\alpha E_\alpha \sin\theta_\alpha\ ,
\label{Et}
\eeq
while the missing transverse energy $\met$ was already defined in (\ref{met}).

We are now in a position to introduce the variable $H_T$
which is commonly used throughout the literature, yet, quite
surprisingly, there is no universally accepted definition for it.
The idea behind $H_T$ is to add up the transverse energies of
various objects in the event, {\em including} the missing 
energy (\ref{met}). While the idea is rather straightforward, 
there are large variations when it comes to its implementation.
For example, one issue is whether one should use only 
reconstructed objects or simply sum over all calorimeter towers 
as we have been doing here so far. The former method has the advantage that it 
would tend to reduce pollution from the underlying event, noise, etc.
On the other hand, it would introduce dependence on the 
jet reconstruction algorithm, the ID cuts, etc.
Those subtleties are avoided in the second method, which 
defines a purely calorimeter based $H_T$. There are other possible 
variations in the definition of $H_T$, for example, whether one
includes all jets, or just the top 4 in $p_T$ \cite{Tovey:2000wk}, 
whether or not one includes the leptons in the sum, etc. 
For the purposes of this paper, we do not need to go 
into such details, and we shall simply use a calorimeter-based, 
all inclusive $H_T$ definition as
\beq
H_T \equiv E_T + \met \ .
\label{Ht}
\eeq
Finally, we shall also consider the total visible mass in the event
\cite{Hubisz:2008gg}
\beq
M \equiv \sqrt{E^2-P_x^2-P_y^2-P_z^2} = \sqrt{ E^2 - \mpt^2 - P_z^2}\ .
\label{Mvis}
\eeq
Note that in terms of the visible mass $M$ just introduced,
our $\hat{s}^{1/2}_{min}$ variable (\ref{smin_def_met}) 
can be alternatively written in a more symmetric form as
\beq
\hat{s}^{1/2}_{min}(M_{inv}) = \sqrt{\met^2 + M^2}+\sqrt{\met^2 + M_{inv}^2}\ .
\eeq

We are now ready to contrast the so defined global inclusive variables
$E$, $\met$, $E_T$, $H_T$ and $M$ to our variable 
$\hat{s}^{1/2}_{min}$ defined in (\ref{smin_def_met}).
%which can be equivalently rewritten in terms of the missing energy $\met$ as
%\beq
%\hat{s}^{1/2}_{min}(M_{inv}) = \sqrt{E^2-P_z^2}+\sqrt{\met^2+M_{inv}^2}\ .
%\label{smin_def_met}
%\eeq
Since $\hat{s}^{1/2}_{min}(M_{inv})$ depends on the a priori unknown 
invisible mass parameter $M_{inv}$, first we need to decide what to do
about the $M_{inv}$ dependence in (\ref{smin_def_met}). In the remainder 
of this section, we shall adopt a most conservative approach: we will
simply set $M_{inv}=0$ and consider the variable 
\beq
\hat{s}^{1/2}_{min}(0) = \sqrt{E^2-P_z^2}+\met\ .
\label{smin_def_met0}
\eeq
This choice is indeed very conservative: for SM processes, where the 
missing energy is due to neutrinos, this would be the proper variable 
to use anyway. On the other hand, for BSM processes with massive 
invisible particles, at this point we are lacking the necessary information
to make a more informed choice. We shall postpone our quantitative 
discussion of the $M_{inv}$ dependence in (\ref{smin_def_met}) until
the next section \ref{sec:mass}.

\FIGURE[t]{
\epsfig{file=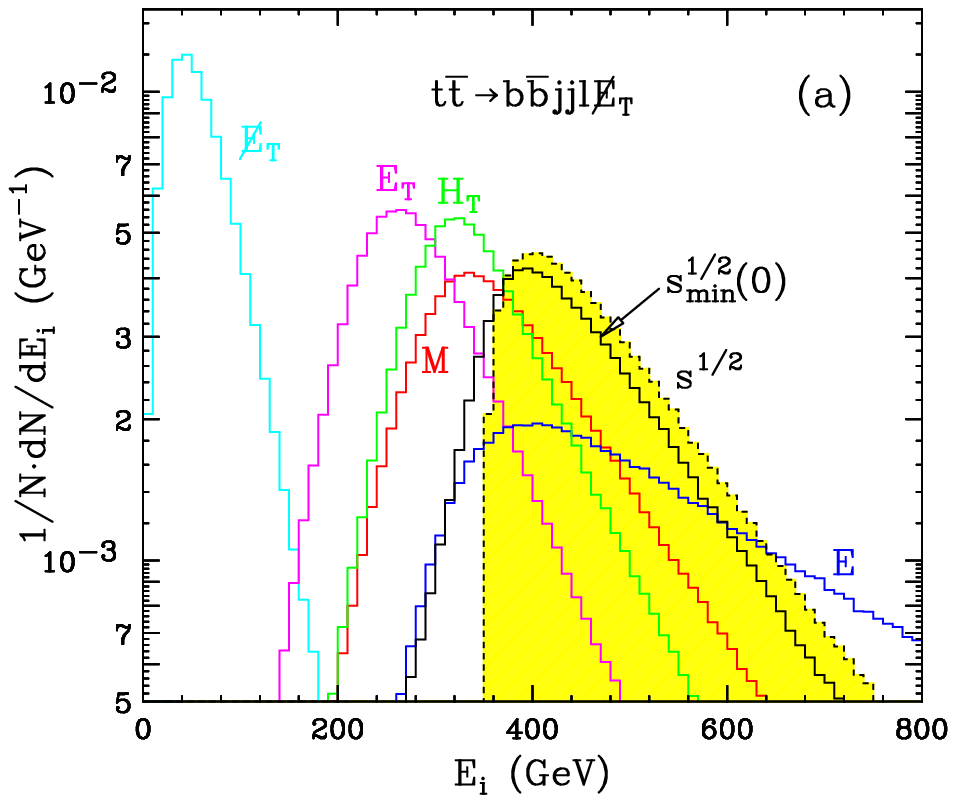,width=7.0cm}~~~
\epsfig{file=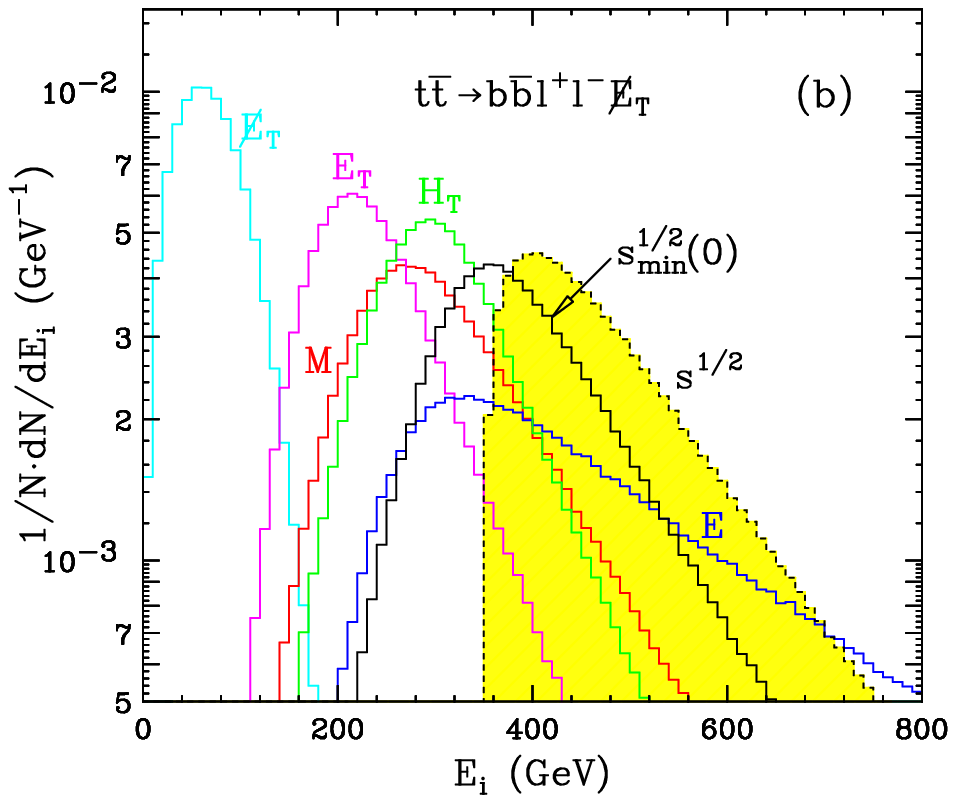,width=7.0cm}
\caption{\sl Unit-normalized distributions of the various energy scale variables
$E_i$ introduced in Section~\ref{sec:compare}:
$E$ (blue), $\met$ (cyan), $E_T$ (magenta), $H_T$ (green), $M$ (red)
and $\hat{s}^{1/2}_{min}(0)$ (black); in (a) single-lepton and (b) dilepton 
$t\bar{t}$ events.
The dotted (yellow-shaded) histograms are identical in panels (a) and (b)
and show the true $\hat{s}^{1/2}$ distribution.
}
\label{fig:varstt}}

We shall illustrate our comparisons with specific examples, illustrated in
Figs.~\ref{fig:varstt}, \ref{fig:varsgg} and \ref{fig:varscg}.
In each case, we shall plot the six different global inclusive variables 
$E_i$ introduced so far, with the following color scheme: in
Figs.~\ref{fig:varstt}-\ref{fig:varscg} we shall plot
the calorimeter energy $E$ (\ref{E}) with blue lines,
the missing transverse energy $\met$ (\ref{met}) with cyan lines,
the total transverse energy $E_T$ (\ref{Et}) with magenta lines,
the $H_T$ variable (\ref{Ht}) with green lines, 
the total visible mass $M$ (\ref{Mvis}) with red lines,
and finally, our $\hat{s}^{1/2}_{min}(0)$ variable (\ref{smin_def_met0})
with solid black lines. All numerical results shown here have been
obtained with PYTHIA\footnote{For simplicity, for the numerical results 
shown in this and the next two sections, we turned off ISR and MPI in PYTHIA, 
which allows us to better illustrate and subsequently explain the 
salient features of $\hat{s}^{1/2}_{min}$. The ISR and MPI effects 
will be studied later in Section~\ref{sec:ISR}.} 
\cite{Sjostrand:2006za} and the PGS detector simulation package \cite{PGS}.
As our first example, shown in Fig.~\ref{fig:varstt}, 
we choose $t\bar{t}$ production at the LHC (the corresponding data 
from the Tevatron already exists, so the same comparison can also
be made directly with CDF and D0 data as well). In Fig.~\ref{fig:varstt}(a)
(Fig.~\ref{fig:varstt}(b)) we show our results for the semi-leptonic (dilepton) channel.
The dilepton $t\bar{t}$ sample is rather similar to a hypothetical new physics signal due to
dark matter particle production: each event has a certain amount of missing
energy, which is due to {\em two} invisible particles escaping the detector.

In each panel of Fig.~\ref{fig:varstt}, the dotted (yellow-shaded) histogram
shows the true $\hat{s}^{1/2}$ distribution, which is the one we would
ideally want to measure. However, due to the missing neutrinos, 
$\hat{s}^{1/2}$ is not directly observable, unless we make some further 
assumptions and attempt some kinematical event reconstruction. 
Therefore we concentrate on the remaining distributions shown in Fig.~\ref{fig:varstt}, 
which are immediately and directly observable. In particular, we shall
pose the question, which among the various distributions exhibited
in Fig.~\ref{fig:varstt} seems to be the best approximation to the true
$\hat{s}^{1/2}$ distribution. A quick glance at Fig.~\ref{fig:varstt} 
reveals that the variable which comes closest to the true $\hat{s}^{1/2}$ is
precisely our variable $\hat{s}^{1/2}_{min}(0)$ defined in (\ref{smin_def_met0}).
As for the rest, we see that the missing transverse energy $\met$ is a very 
poor estimator of the energy scale of the events, while $E_T$, $H_T$ and $M$ 
are doing a little bit better, yet are still quite far off.
As can be expected from its definition 
(\ref{Ht}), $H_T$ is always somewhat larger than $E_T$, while $H_T$ and $M$
are rather similar, with $H_T$ ($M$) doing better for the dilepton 
(semi-leptonic) case. Finally, the total energy $E$ is relatively 
close to the true $\hat{s}^{1/2}$ distribution, but is quite broad
in both Figs.~\ref{fig:varstt}(a) and \ref{fig:varstt}(b). In contrast,
the $\hat{s}^{1/2}_{min}(0)$ distribution is quite sharp, and 
is thus a better indicator of the relevant energy scale.

Let us now take a closer look at the two $\hat{s}^{1/2}$ distributions in each
panel of Fig.~\ref{fig:varstt}. Since $\hat{s}^{1/2}_{min}$ was defined through
a minimization procedure, it is clear that it will always underestimate the true
$\hat{s}^{1/2}$. Fig.~\ref{fig:varstt} quantifies the amount of this 
underestimation for the case of $t\bar{t}$ events. We see that 
$\hat{s}^{1/2}_{min}(0)$ is tracking the true $\hat{s}^{1/2}$
quite well for the case of semi-leptonic $t\bar{t}$ events in Fig.~\ref{fig:varstt}(a). 
This could have been expected on very general grounds: for semi-leptonic events, 
we are missing a single neutrino, whose transverse momentum is actually measured 
through $\mptvec$, so that the only mistake we are making in approximating
$\hat{s}^{1/2}\approx \hat{s}^{1/2}_{min}(0)$ is due to the unknown 
longitudinal component $p_{1z}$. In the case of dilepton events, however, 
there are two missing neutrinos, and thus more unknown degrees of freedom which
we have to fix rather ad hoc according to our prescription (\ref{ptsol}, \ref{pzsol}).
The resulting error is larger and leads to a larger displacement between the 
true $\hat{s}^{1/2}$ distribution and its $\hat{s}^{1/2}_{min}(0)$ approximation,
as can be seen in Fig.~\ref{fig:varstt}(b).

In the case of $t\bar{t}$ illustrated in Fig.~\ref{fig:varstt} the missing 
energy arises from massless SM neutrinos, so that the approximation $M_{inv}=0$
is well justified. Let us now consider a situation where the observed missing energy 
signal is due to {\em massive} neutral stable particles, as opposed to SM neutrinos.
The prototypical example of this sort is low energy supersymmetry with conserved 
$R$-parity, and this is what we shall use for our next two examples as well. 
Each SUSY event will be initiated by the pair-production of two superpartners, 
which will then cascade decay to the lightest supersymmetric particle (LSP), which we
shall assume to be the lightest neutralino $\tilde \chi^0_1$. Since there are 
two SUSY cascades per event, there will be two LSP particles in the final state, 
so that
\beq
n_{inv}=n_\chi=2\ . 
\label{ninvsusy}
\eeq
Furthermore, since the two LSPs are identical, we also have
\beq
m_1=m_2\equiv m_\chi\ ,
\label{mchidef}
\eeq
i.e. in what follows we shall denote the true LSP mass with $m_\chi$.
From (\ref{minv}), (\ref{ninvsusy}) and (\ref{mchidef}) it follows
that the true total invisible mass in any SUSY event is simply
\beq
M_{inv} = 2 m_\chi\ .
\eeq
However, the true LSP mass $m_\chi$ is a priori unknown, 
therefore, when we construct our variable 
\beq
\hat{s}^{1/2}_{min}(M_{inv})=\hat{s}^{1/2}_{min}(2m_\chi)
\eeq
for the SUSY examples, we will have to make a guess for the 
value of the LSP mass $m_\chi$. We shall denote this trial
value by $\tilde m_\chi$, in order to distinguish it from the true
LSP mass $m_\chi$. This situation is reminiscent of the 
case of the Cambridge $M_{T2}$ variable \cite{Lester:1999tx}, 
where in order to construct the $M_{T2}$ variable itself, 
one must first choose a test value for the LSP mass.
Our notation here is consistent with the notation for $M_{T2}$
used in \cite{Burns:2008va}.

\FIGURE[t]{
\epsfig{file=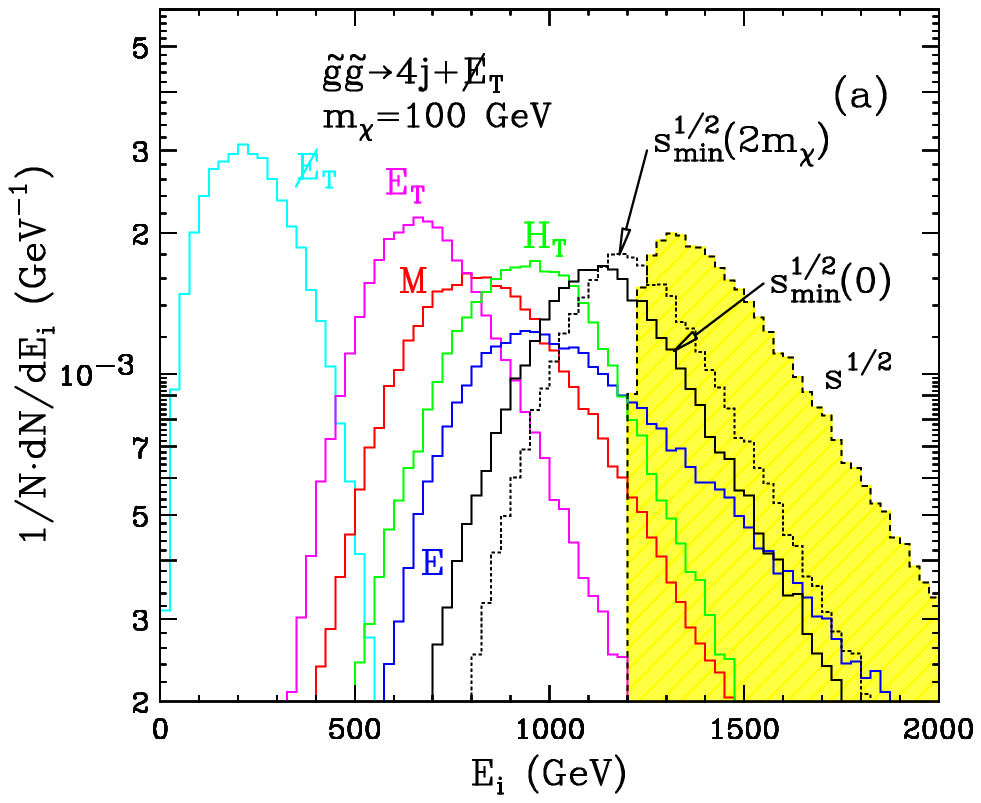,width=7.0cm}~~~
\epsfig{file=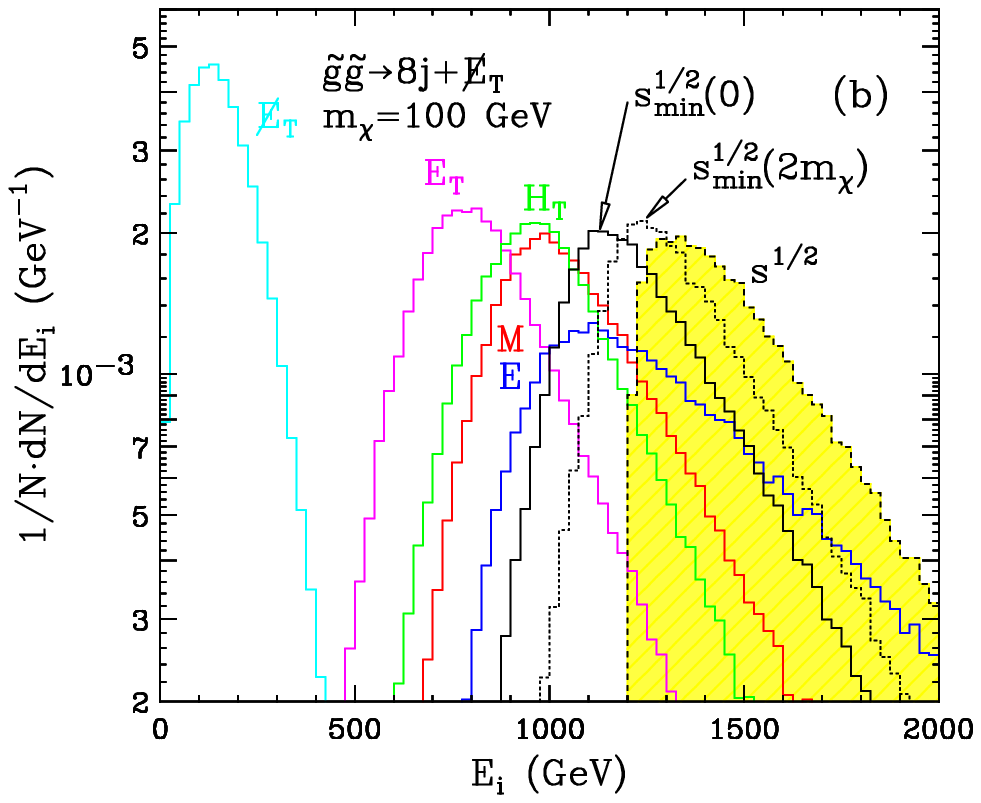,width=7.0cm}
\caption{\sl The same as Fig.~\ref{fig:varstt}, but for gluino pair 
production events with (a) 2-jet gluino decays as in (\ref{2jet})
and (b) 4-jet gluino decays as in (\ref{4jet}).
The SUSY masses are fixed as follows: 
$m_{\tilde \chi^0_1}=100$ GeV, $m_{\tilde \chi^0_2}=200$ GeV and
$m_{\tilde g}=600$ GeV.
In addition to the variables shown in Fig.~\ref{fig:varstt},
here we also plot the $\hat{s}^{1/2}_{min}(2m_\chi)$ distribution (dotted line)
with the correct value of the invisible mass $M_{inv}=2m_\chi=2m_{\tilde\chi^0_1}$.
}
\label{fig:varsgg}}

We are now ready to describe our SUSY examples. For our study 
we will choose a rather difficult signature --- jets plus $\met$,
for which all other proposed methods for mass determination
are bound to face significant challenges. For concreteness, 
we consider gluino production, followed by a gluino 
decay to jets and a neutralino. In Fig.~\ref{fig:varsgg}
we consider gluino pair-production ($\tilde g\tilde g$), 
while in Fig.~\ref{fig:varscg} we show results for associated 
gluino-LSP production ($\tilde g\tilde \chi^0_1$).
In addition, we consider two different possibilities for 
the gluino decays. The first case, shown in Figs.~\ref{fig:varsgg}(a)
and Figs.~\ref{fig:varscg}(a), has the gluino decaying 
directly to the LSP:
\beq
\tilde g \to jj\tilde\chi^0_1\ ,
\label{2jet}
\eeq
so that the gluino pair-production events in Fig.~\ref{fig:varsgg}(a)
have 4 jets and missing energy, while the associated gluino-LSP 
production events in Fig.~\ref{fig:varscg}(a) have two jets and 
missing energy. In the second case, presented in Figs.~\ref{fig:varsgg}(b)
and Figs.~\ref{fig:varscg}(b), we forced the gluino
to always decay to $\tilde\chi^0_2$, which in turn decays 
via a 3-body decay to 2 jets and the LSP:
\beq
\tilde g \to jj\tilde\chi^0_2\to jjjj\tilde\chi^0_1\ .
\label{4jet}
\eeq
As a result, the gluino pair-production events in Fig.~\ref{fig:varsgg}(b)
will exhibit 8 jets and missing energy, while the associated gluino-LSP 
production events in Fig.~\ref{fig:varscg}(b) will have four jets and 
missing energy. Of course, the actual number of reconstructed jets in 
such events may be even higher, due to the effects of initial state 
radiation (ISR) and/or jet fragmentation. In any case, such multijet 
events will be very challenging for any exclusive reconstruction method,
therefore it is interesting to see what we can learn about them from
the global inclusive variables discussed here. 

\FIGURE[t]{
\epsfig{file=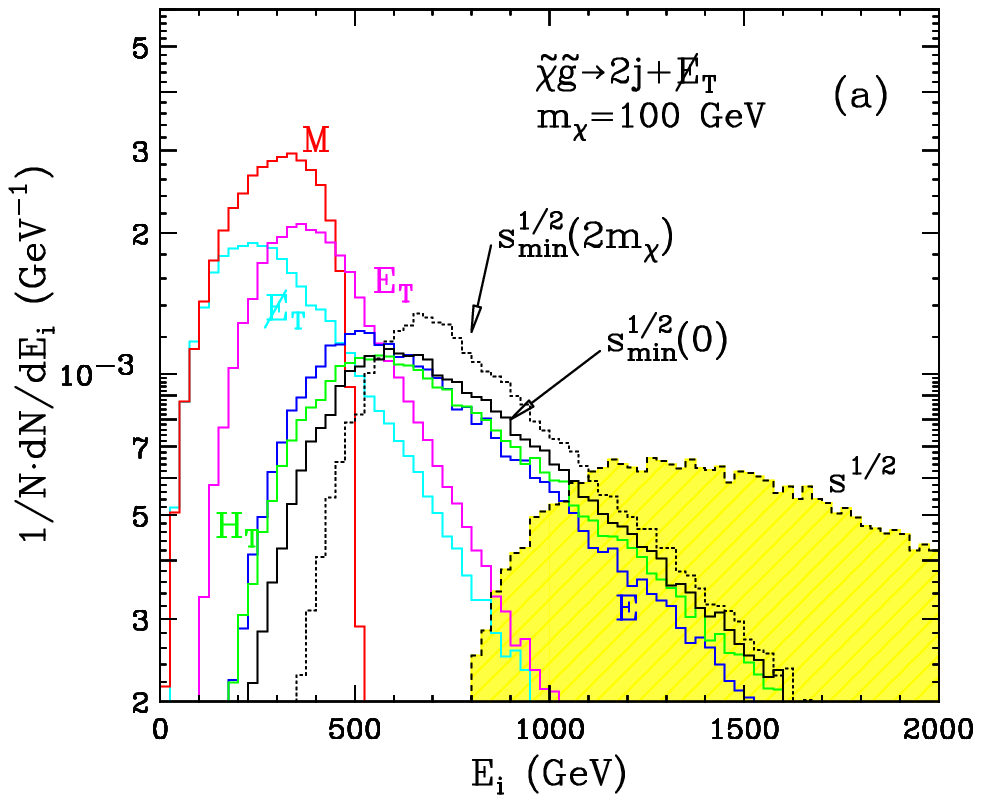,width=7.0cm}~~~
\epsfig{file=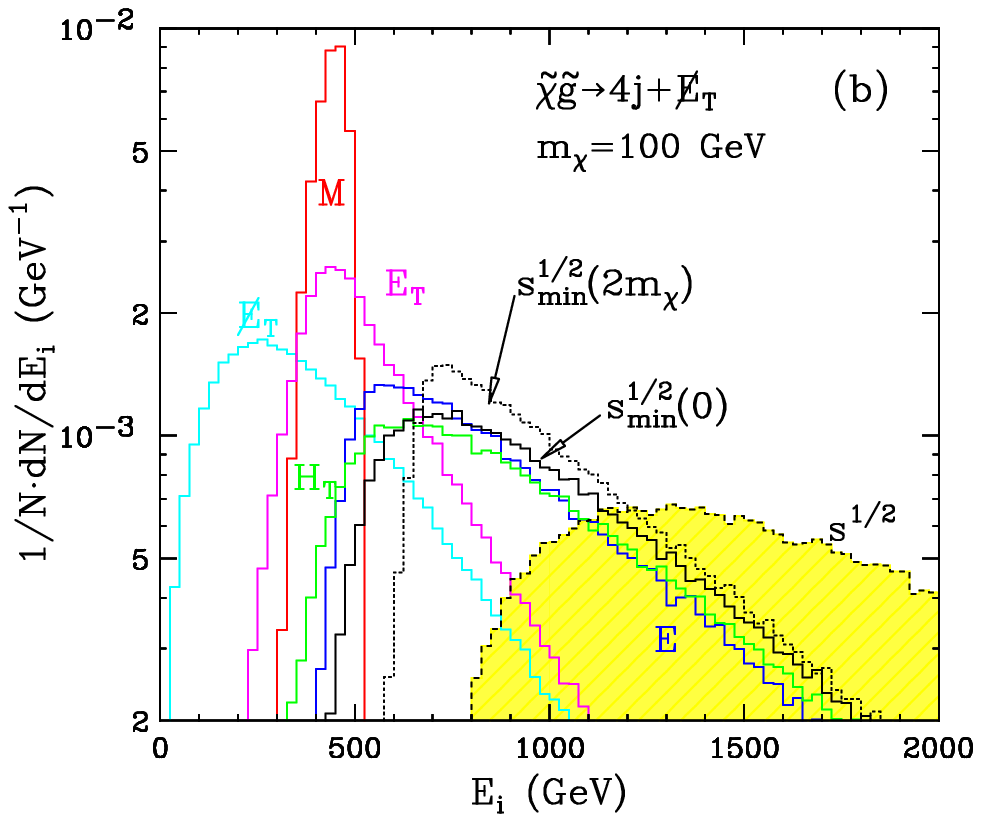,width=7.0cm}
\caption{\sl The same as Fig.~\ref{fig:varsgg}, but for 
events of associated gluino-LSP production.
}
\label{fig:varscg}}

For concreteness, in what follows we shall always fix the relevant 
SUSY masses according to the approximate gaugino unification relation
\beq
m_{\tilde g} = 3 m_{\tilde\chi^0_2} = 6 m_{\tilde\chi^0_1} \equiv 6 m_\chi\ ,
\label{gunif}
\eeq
and since we assume three-body decays in (\ref{2jet}) and (\ref{4jet}),
we do not need to specify the SUSY scalar mass parameters, which can be
taken to be very large. In addition, as implied by (\ref{gunif}),
we imagine that the lightest two neutralinos are gaugino-like, so that
we do not have to specify the higgsino mass parameter either, and it can 
be taken to be very large as well. 

Fig.~\ref{fig:varsgg} shows our results for the different global 
inclusive variables introduced earlier, for the case of gluino 
pair-production. All in all, the outcome is not too different from 
what we found previously in Fig.~\ref{fig:varstt} for the $t\bar{t}$ case:
when it comes to approximating the true $\hat{s}^{1/2}$ distribution,
the missing energy $\met$ does the worst, our variable 
$\hat{s}^{1/2}_{min}(0)$ does the best, and all other remaining 
variables are somewhere in between those two extremes. 
This time, in Fig.~\ref{fig:varsgg} we also plot one ``cheater''
distribution, namely $\hat{s}^{1/2}_{min}(2m_\chi)$, where
we have used the correct value of the invisible mass 
$M_{inv}=2m_\chi=2m_{\tilde\chi^0_1}$. It demonstrates that
knowing the actual value of the LSP mass helps (since
$\hat{s}^{1/2}_{min}(2m_\chi)$ gets closer to the truth),
but is not crucial: the quantity $\hat{s}^{1/2}_{min}(0)$
still does surprisingly well in approximating the true $\hat{s}^{1/2}$.

Notice that when the missing energy in the data is due to 
massive BSM particles, there are two sources of error 
in approximating $\hat{s}^{1/2}\approx \hat{s}^{1/2}_{min}(0)$,
each leading to an underestimation.
By comparing the three different types of $\hat{s}^{1/2}$
distributions shown in each panel of Fig.~\ref{fig:varsgg},
one can see quantitatively the effect of each source.
First, when we take the {\em minimum} possible value 
of $\hat{s}^{1/2}$ in (\ref{shat2}), we are underestimating by a certain amount, 
which can be seen by comparing the ``cheater'' distribution
$\hat{s}^{1/2}_{min}(2m_\chi)$ (dotted line) 
to the $\hat{s}^{1/2}$ truth (yellow shaded). Second,
as we do not know a priori the LSP mass, we take conservatively
$M_{inv}=0$, which leads to a further underestimation, 
as evidenced by the difference between the 
$\hat{s}^{1/2}_{min}(0)$ distribution (solid line) and its ``cheater''
version $\hat{s}^{1/2}_{min}(2m_\chi)$. In spite of those two 
undesirable effects, the $\hat{s}^{1/2}_{min}(0)$ approximation 
that we end up with is still surprisingly close to the real one,
and is certainly the best approximation among the variables
we are considering.

The common thread in our first two examples shown in 
Figs.~\ref{fig:varstt} and \ref{fig:varsgg} was that the 
events were symmetric, i.e. we produce the same type of particles, 
which then decay identically on each side of the event.
As our last example, we shall consider an extreme version 
of an asymmetric event, namely one where all visible particles come 
from the same side of the event, i.e. from a single 
decay chain. The process of associated gluino-LSP 
production is exactly of this type - all 
jets arise from the decay chain of a single gluino, which is
recoiling against an LSP. The topology of these events 
is very different from the events considered earlier in 
Figs.~\ref{fig:varstt} and \ref{fig:varsgg}. Nevertheless,
as seen in Fig.~\ref{fig:varscg}, we find very similar results.
In particular, among all the different global inclusive variables 
that we are considering, the quantity $\hat{s}^{1/2}_{min}(0)$ is still 
the one closest to the true $\hat{s}^{1/2}$ distribution.

\section{Dependence of $\hat{s}^{1/2}_{min}$ on the unknown masses of invisible particles}
\label{sec:mass}

In the previous Section \ref{sec:compare} we demonstrated the advantage of 
$\hat{s}^{1/2}_{min}$ in comparison to the other commonly used global 
inclusive event variables. From now on we shall therefore focus
our discussion entirely on $\hat{s}^{1/2}_{min}$ and its properties.
In this Section we shall investigate in more detail the dependence  
of $\hat{s}^{1/2}_{min}$ on the (a priori unknown) masses of the invisible 
particles which are causing the observed missing energy signal.
Then in the next Section~\ref{sec:thr} we shall use these results
to correlate the observed $\hat{s}^{1/2}_{min}$ distribution to the
masses of the {\em parent} particles which were
originally produced in the event.

Recall that in the three examples from the previous section, 
we always conservatively chose the invisible mass to be zero:
$M_{inv}=0$ and we correspondingly considered $\hat{s}^{1/2}_{min}(0)$.
This choice is actually a good starting point in studying any missing 
energy signature by means of $\hat{s}^{1/2}_{min}(M_{inv})$.
The assumption of $M_{inv}=0$ is precisely what one would do 
if one were to assume that 
the missing energy is simply due to SM neutrinos, as opposed to 
some new physics. However, if the observed missing energy signal 
is in excess of the expected SM backgrounds, then an alternative,
BSM explanation for those events must be sought. In that case,
we would not know the mass of the invisible particles,
and we would have to make a guess. Our main goal in this section
is to study numerically the effect of this guess.
Our philosophy will be to revisit the SUSY
examples from Section~\ref{sec:compare} and simply vary the 
test mass $\tilde m_\chi$ of the invisible particles (the LSPs). 
Since the two LSPs are identical (see eq.~(\ref{mchidef})), 
we will take their test masses to be the same as well. 

\FIGURE[t]{
\epsfig{file=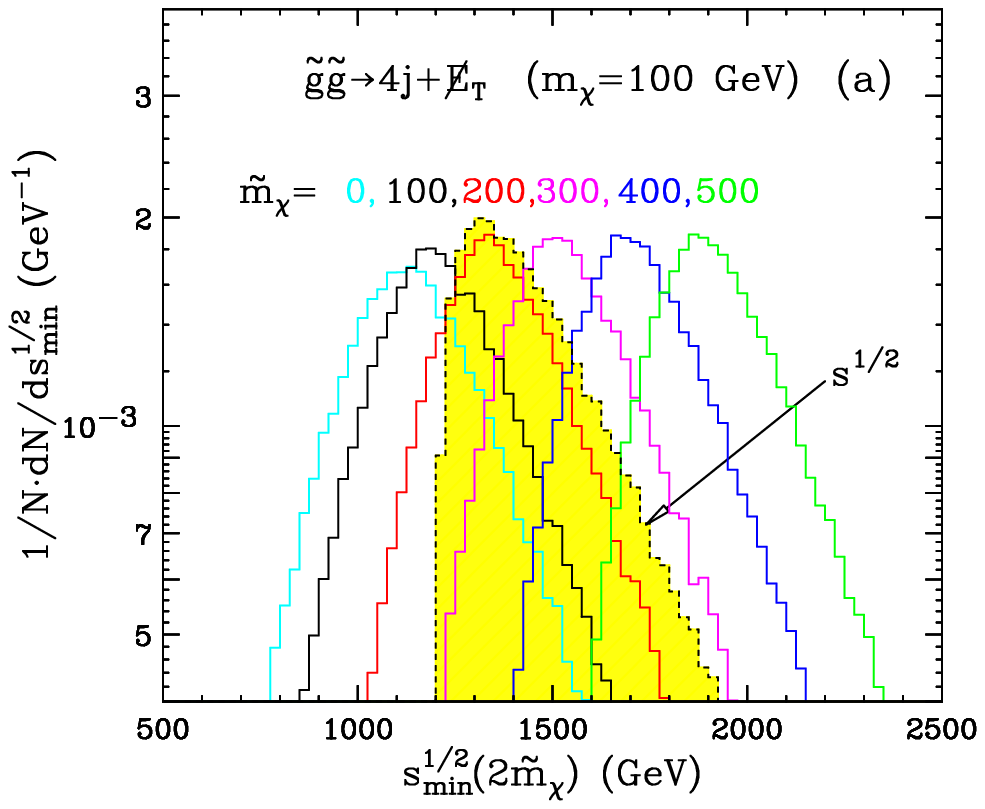,width=7.0cm}~~~
\epsfig{file=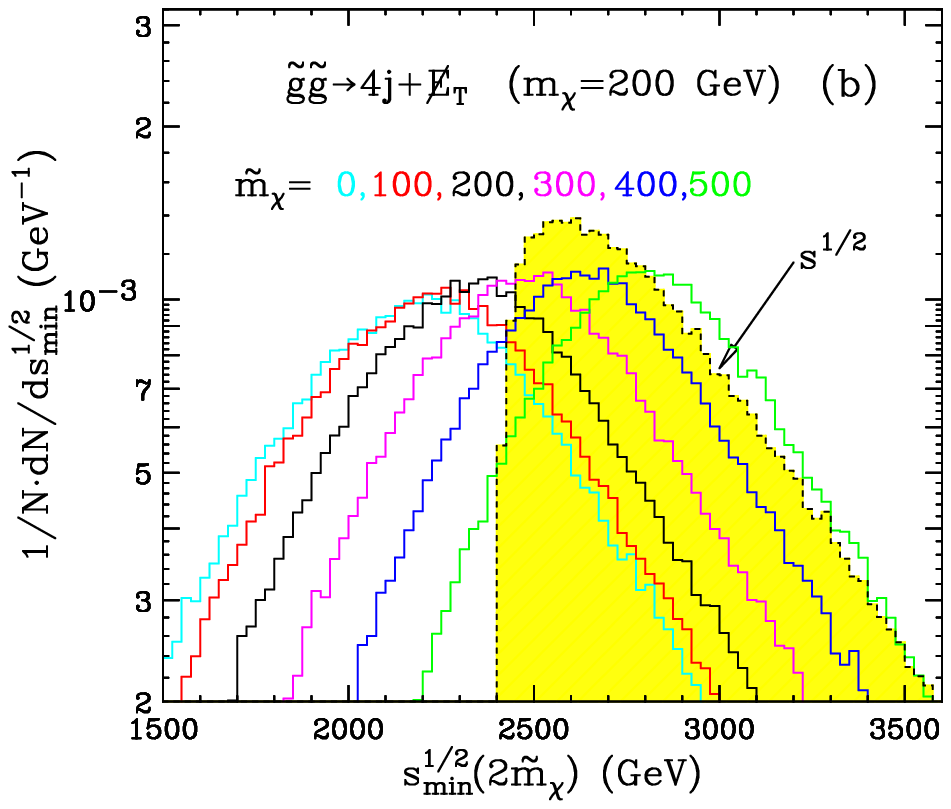,width=6.8cm}\\
\\
\epsfig{file=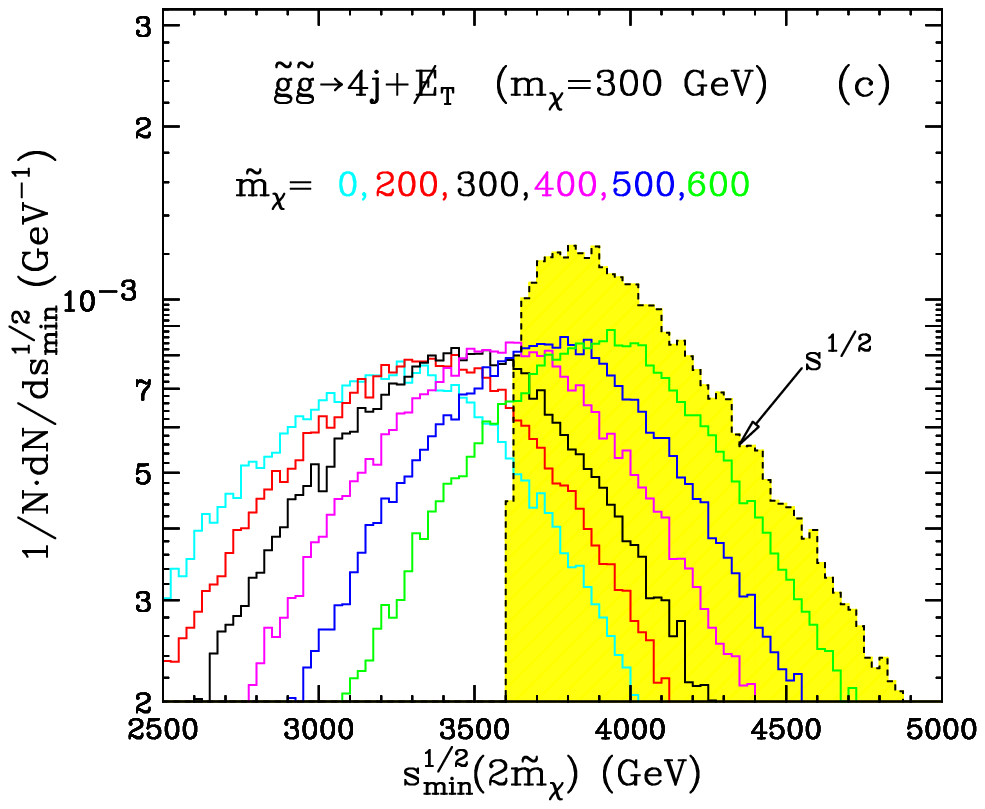,width=7.0cm}~~~
\epsfig{file=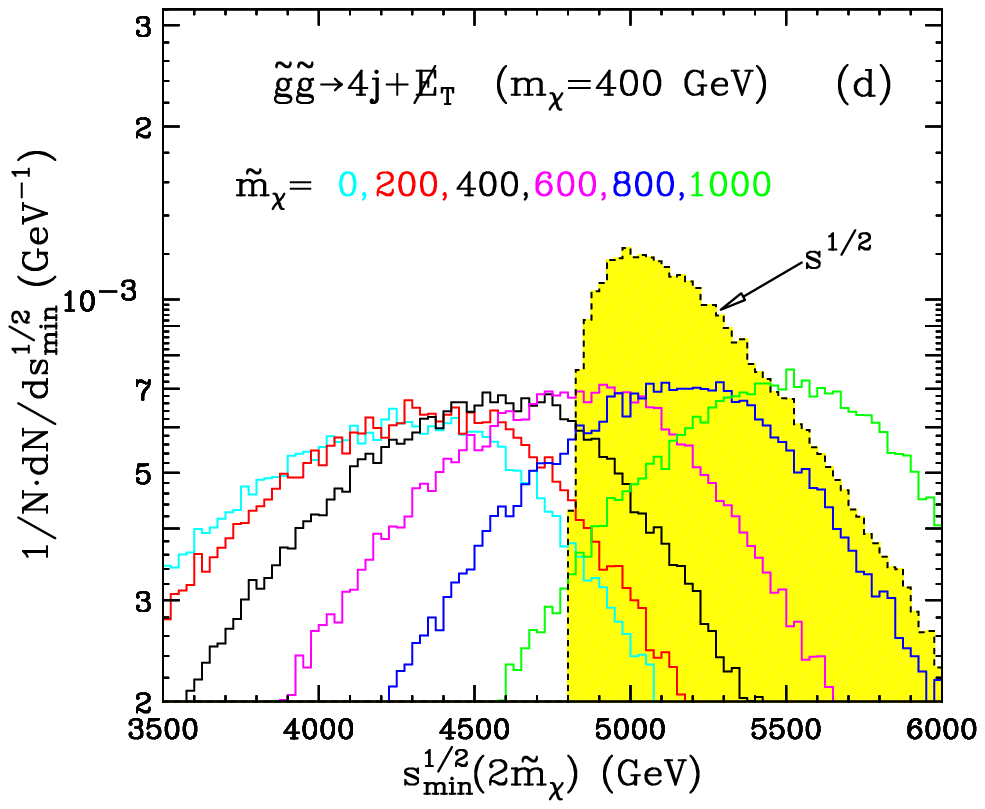,width=7.0cm}

\caption{\sl Unit-normalized distributions of the $\hat{s}^{1/2}_{min}(M_{inv})$ 
variable for several different SUSY mass spectra: 
(a) $m_{\tilde\chi^0_1}=100$ GeV, (b) $m_{\tilde\chi^0_1}=200$ GeV,
(c) $m_{\tilde\chi^0_1}=300$ GeV, and (d) $m_{\tilde\chi^0_1}=400$ GeV.
The remaining masses are fixed according to (\ref{gunif}).
We consider gluino pair-production events with 2-jet gluino 
decays as in (\ref{2jet}). In each panel, we plot the 
$\hat{s}^{1/2}_{min}(M_{inv})=\hat{s}^{1/2}_{min}(2\tilde m_\chi)$ 
distributions for several representative values of the trial LSP mass 
$\tilde m_\chi$ as shown. The color scheme is such that the black histogram
is always the case where we happen to use the correct value 
of the LSP mass: $\tilde m_\chi = m_\chi$. 
The dotted (yellow-shaded) histogram gives the true 
$\hat{s}^{1/2}$ distribution.}
\label{fig:gg2j}}

Our results are presented in Figs.~\ref{fig:gg2j}, \ref{fig:gg4j} and \ref{fig:cg4j}.
In Figs.~\ref{fig:gg2j} and \ref{fig:gg4j}
we consider gluino pair production. In Fig.~\ref{fig:gg2j}
each gluino decays to 2 jets as in (\ref{2jet}), while
in Fig.~\ref{fig:gg4j} each gluino decays to 4 jets as in (\ref{4jet}). 
Then in Fig.~\ref{fig:cg4j} we consider asymmetric events
of associated gluino-LSP production, where the single gluino decays
to 4 jets as in (\ref{4jet}). In each figure, we consider 
four different study points, defined through the value of the true
LSP mass $m_\chi$. In all three Figs.~\ref{fig:gg2j}-\ref{fig:cg4j},
panels (a) correspond to $m_\chi=100$ GeV, 
panels (b) have $m_\chi=200$ GeV, panels (c) have $m_\chi=300$ GeV, 
while in panels (d) $m_\chi=400$ GeV. As before, the remaining masses 
$m_{\tilde g}$ and $m_{\tilde\chi^0_2}$ are always fixed 
according to the approximate gaugino unification relation
(\ref{gunif}). Each panel in Figs.~\ref{fig:gg2j}-\ref{fig:cg4j}
exhibits the true $\hat{s}^{1/2}$ distribution (yellow-shaded histogram), 
and the corresponding $\hat{s}^{1/2}_{min}(2\tilde m_\chi)$
distributions for several representative values of the test LSP mass
$\tilde m_\chi$. Each $\hat{s}^{1/2}_{min}$ curve is both 
color coded and labelled by its corresponding value of $\tilde m_\chi$.
Our color scheme is such that the $\hat{s}^{1/2}_{min}$ histogram 
in black is the one where we happen to use the correct value of the 
LSP mass, i.e. when $\tilde m_\chi=m_\chi$.

\FIGURE[t]{
\epsfig{file=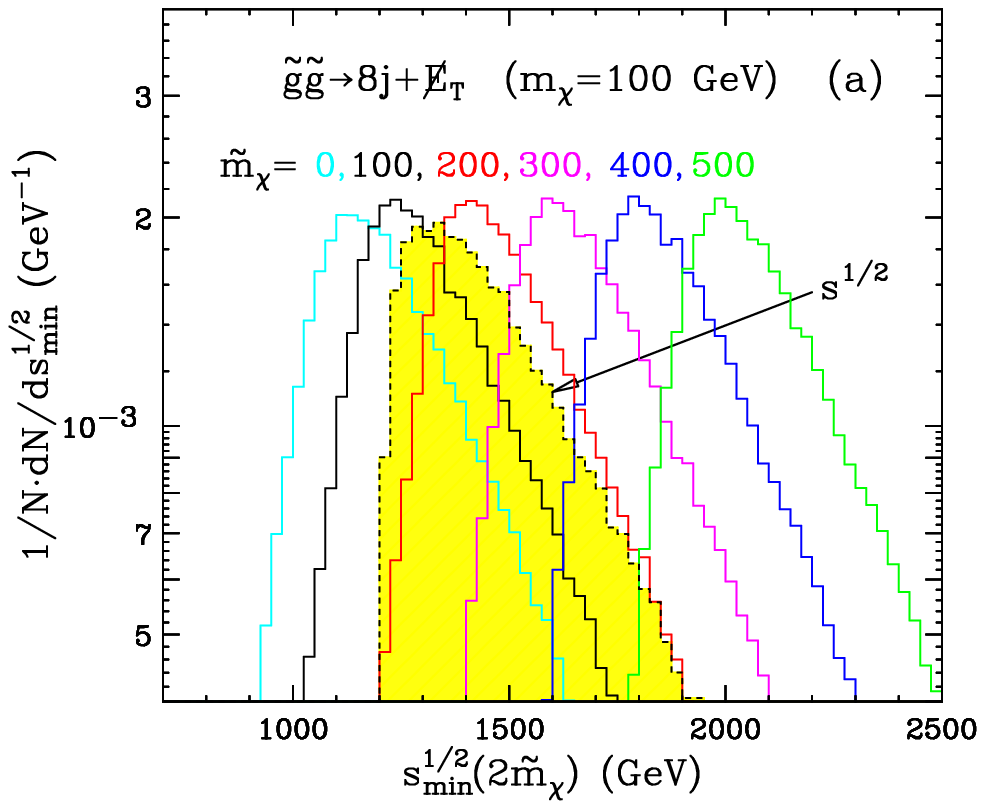,width=7.0cm}~~~
\epsfig{file=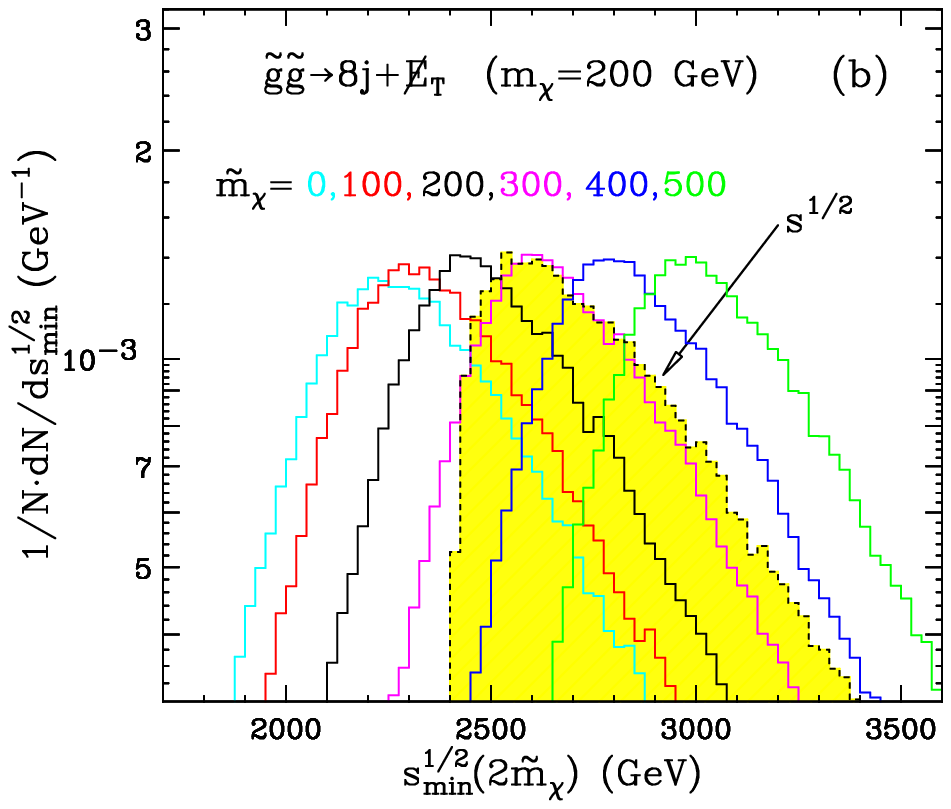,width=6.8cm}\\
\\
\epsfig{file=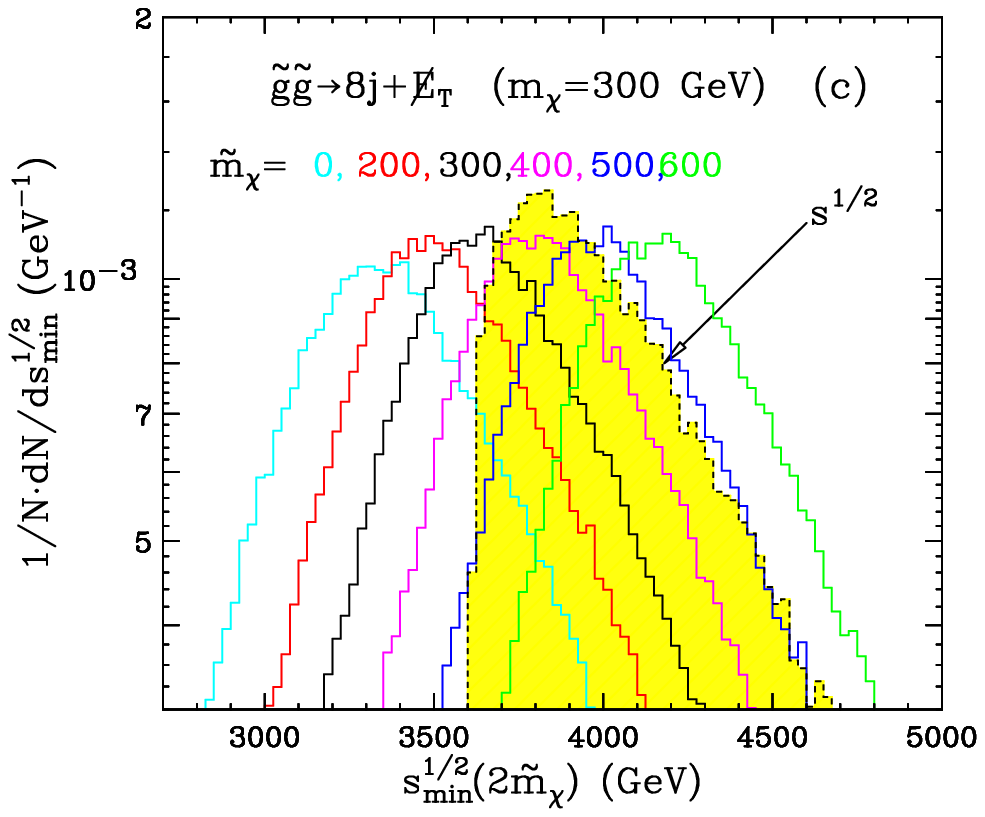,width=7.0cm}~~~
\epsfig{file=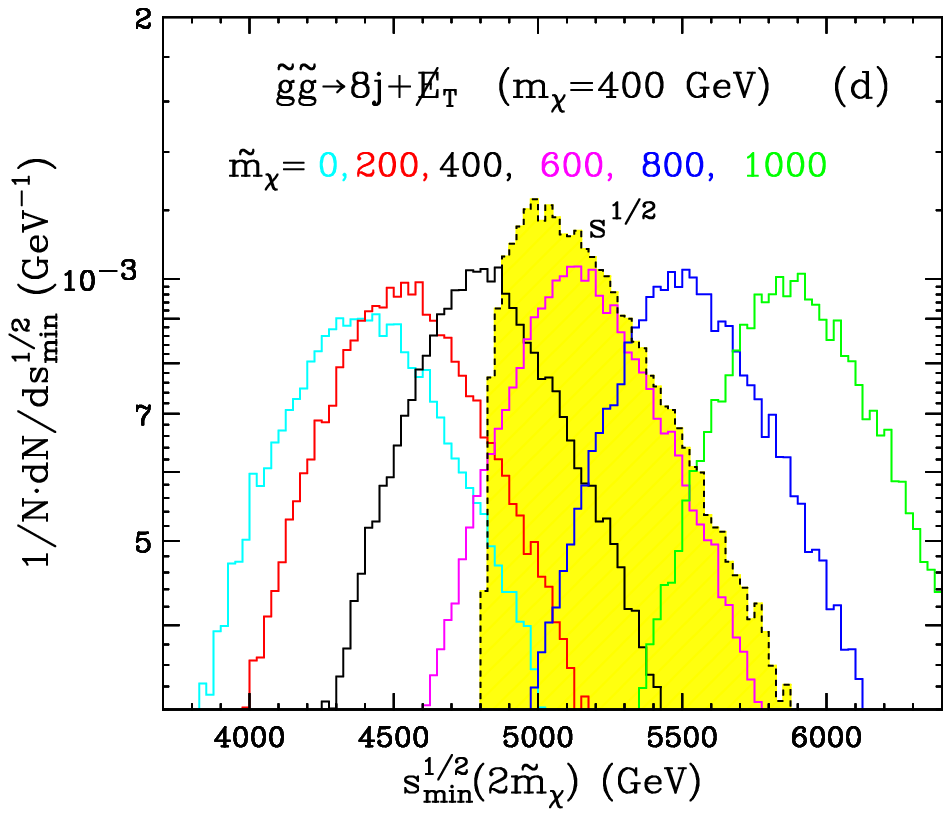,width=7.0cm}
\caption{\sl The same as Fig.~\ref{fig:gg2j}, but for 
4-jet gluino decays as in (\ref{4jet}).
}
\label{fig:gg4j}}

The qualitative behavior seen in Figs.~\ref{fig:gg2j}-\ref{fig:cg4j}
is more or less as expected: the $\hat{s}^{1/2}_{min}(2\tilde m_\chi)$
distributions shift to higher energy scales, as we increase the 
value of the test mass $\tilde m_\chi$. This can be easily understood
from the definition (\ref{smin_def_met}) of the $\hat{s}^{1/2}_{min}(M_{inv})$ 
variable: for any given set of $E$, $P_z$ and $\met$ values, 
$\hat{s}^{1/2}_{min}(M_{inv})$ is a monotonically increasing
function of $M_{inv}$. The shifts observed in Figs.~\ref{fig:gg2j}-\ref{fig:cg4j}
also make perfect physical sense: obviously, one needs more energy in order to 
produce heavier invisible particles.

\FIGURE[t]{
\epsfig{file=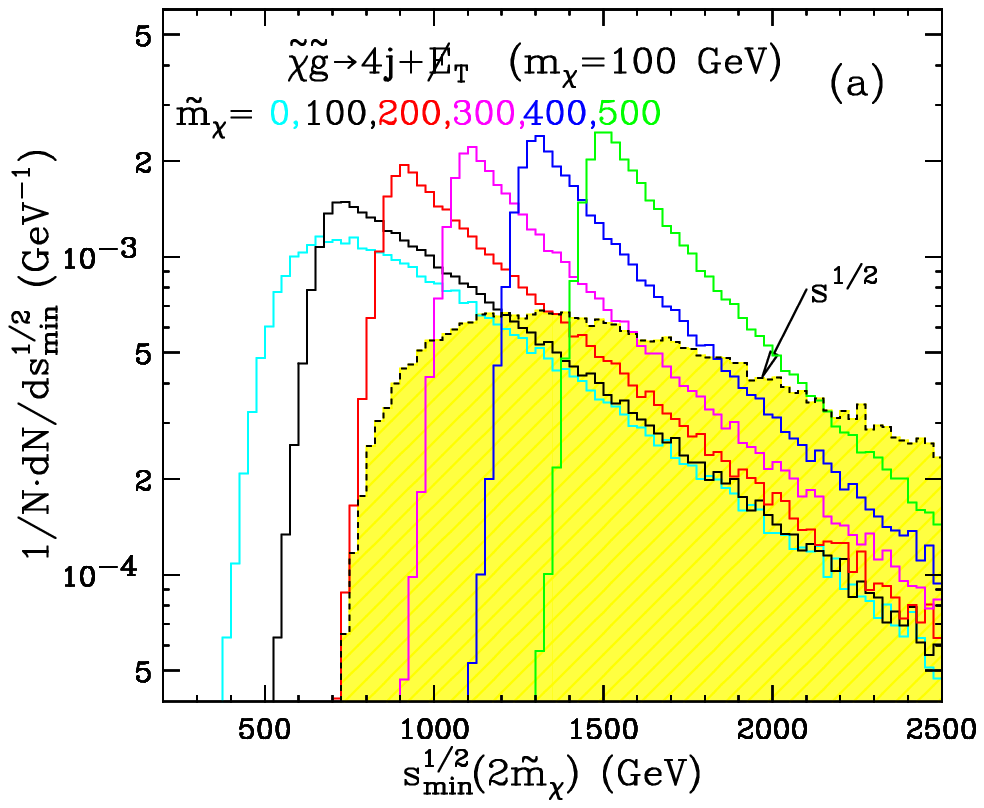,width=7.0cm}~~~
\epsfig{file=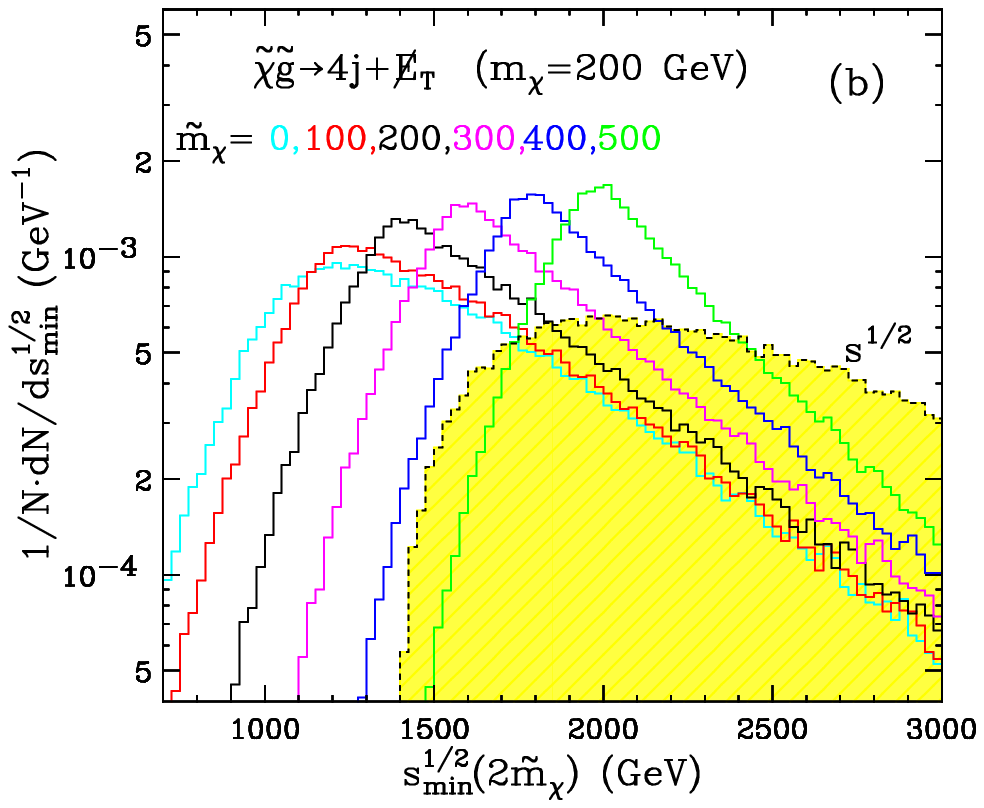,width=7.0cm}\\
\\
\epsfig{file=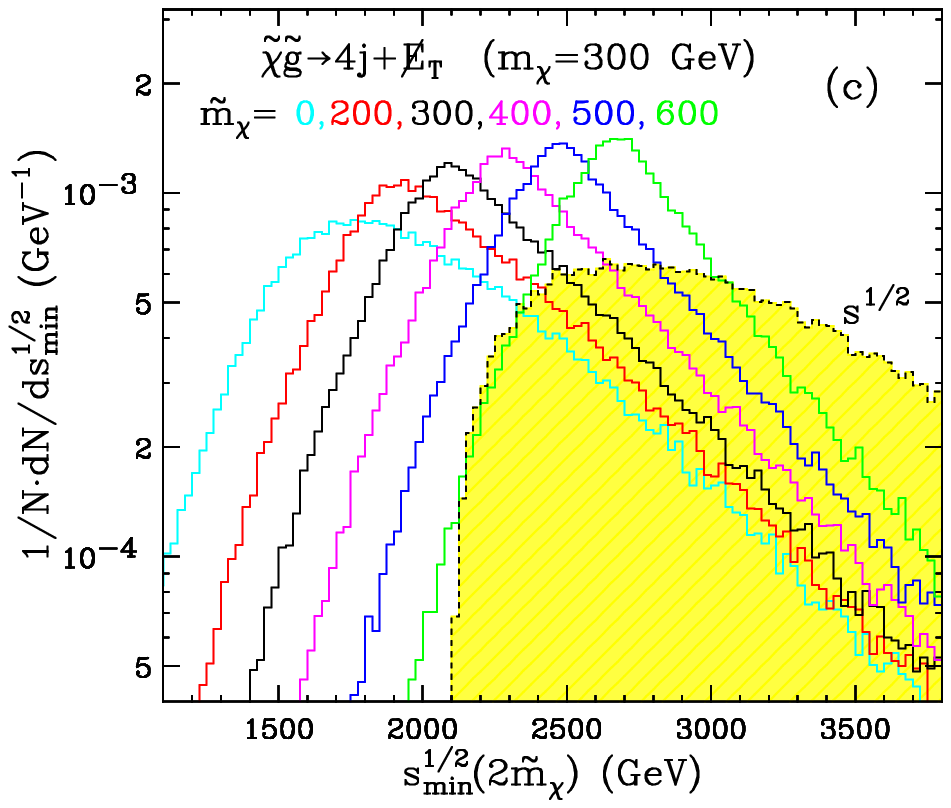,width=7.0cm}~~~
\epsfig{file=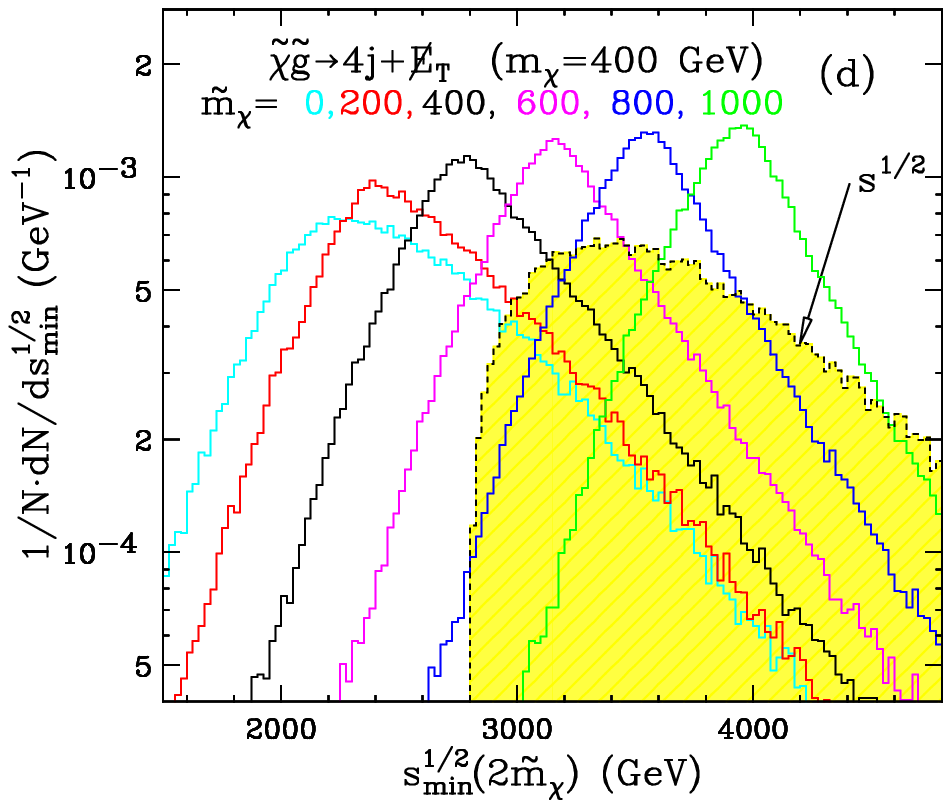,width=7.0cm}
\caption{\sl The same as Fig.~\ref{fig:gg4j}, but for 
events of associated gluino-LSP production ($\tilde g\tilde\chi^0_1$).
}
\label{fig:cg4j}}

Let us now concentrate on the quantitative aspects of 
Figs.~\ref{fig:gg2j}-\ref{fig:cg4j}. Upon careful inspection of 
the three figures, we notice that when the test mass $\tilde m_\chi$
is equal to the true mass $m_\chi$ (i.e. for the black colored histograms),
the corresponding distribution $\hat{s}^{1/2}_{min}(2m_\chi)$
peaks very close to the true $\hat{s}^{1/2}$ threshold 
$\left(\hat{s}^{1/2}\right)_{thr}$. As usual, we define the threshold $\left(\hat{s}^{1/2}\right)_{thr}$
as the value where the true $\hat{s}^{1/2}$ distribution 
(yellow shaded histogram) sharply turns on. This observation
is potentially extremely important, 
since the threshold $\left(\hat{s}^{1/2}\right)_{thr}$ is simply related 
to the masses of the two particles which were originally 
produced in the event. For example, for the gluino 
pair production events in Figs.~\ref{fig:gg2j} and \ref{fig:gg4j}
the threshold is given by
\beq
\left(\hat{s}^{1/2}\right)_{thr} = 2m_{\tilde g} = 12 m_\chi\ ,
\label{sthr_gg}
\eeq
where the second equality is valid only under the
gaugino unification assumption (\ref{gunif}).
Similarly, in the case of associated gluino-LSP production 
in Fig.~\ref{fig:cg4j}, the threshold is given by
\beq
\left(\hat{s}^{1/2}\right)_{thr} = m_{\tilde g} + m_{\tilde\chi^0_1} = 7 m_\chi\ ,
\label{sthr_cg}
\eeq
where once again the second equality is due to our 
assumption (\ref{gunif}). It is easy to verify that 
in all three figures \ref{fig:gg2j}, \ref{fig:gg4j} and \ref{fig:cg4j},
the $\hat{s}^{1/2}$ thresholds (i.e. the sharp turn-ons in the 
yellow-shaded distributions) always occur at the locations predicted in
eqs.~(\ref{sthr_gg}) and (\ref{sthr_cg}). 

Let us now introduce one last piece of notation. In what follows 
we shall use the notation 
\beq
\left(\hat{s}^{1/2}_{min}(M_{inv})\right)_{peak}
\label{speak}
\eeq
to denote the particular value of $\hat{s}^{1/2}_{min}$ where we find 
the peak of the distributions
\beq
\frac{dN(\hat{s}^{1/2}_{min}(M_{inv}))}{d\hat{s}^{1/2}_{min}}
\eeq
which are plotted in Figs.~\ref{fig:gg2j}-\ref{fig:cg4j}.
In other words,
\beq
\left[
\frac{d}{d\hat{s}^{1/2}_{min}}
\frac{dN(\hat{s}^{1/2}_{min}(M_{inv}))}{d\hat{s}^{1/2}_{min}}
\right]_{\hat{s}^{1/2}_{min}=\left(\hat{s}^{1/2}_{min}(M_{inv})\right)_{peak}}
=0.
\label{speakdef}
\eeq
With those conventions, we can now formulate our empirical
observation above as
\beq
\left(\hat{s}^{1/2}\right)_{thr}\approx \left(\hat{s}^{1/2}_{min}(2m_\chi)\right)_{peak} \ .
\label{sminsthr}
\eeq
The last equation is one of the main results in this paper.
While we were not able to derive it in a strict mathematical sense,
it is nevertheless supported by our numerical results shown
in Figs.~\ref{fig:gg2j}-\ref{fig:cg4j}. We also checked many other 
SUSY examples, where we used different mass spectra 
and different production processes and decays. We found that
in all cases the approximate relation (\ref{sminsthr}) still holds.
Fig.~\ref{fig:ds} quantifies this statement for the two previously
considered processes of gluino pair production and associated
gluino-LSP production, where the gluinos are forced to decay 
either to 2 jets as in (\ref{2jet}) or to 4 jets as 
in (\ref{4jet}). In the figure we compare the following three 
quantities, all of which are related in one way or another 
to the energy scale $\hat{s}^{1/2}$ of the events:
\begin{itemize}
\item $\left(\hat{s}^{1/2}\right)_{ave}$ : this is the average 
of the true $\hat{s}^{1/2}$ distribution (the one shown 
in the previous figures with the yellow-shaded histogram). 
Here we had to pick some variable which would characterize 
the true $\hat{s}^{1/2}$ distribution. 
Two alternative choices which we also considered were 
the peak or the mean of the true $\hat{s}^{1/2}$ distribution.
All three of these variables are numerically quite close, with
the peak value typically being the lowest, and the average value 
being the largest. In the end we chose $\left(\hat{s}^{1/2}\right)_{ave}$ 
for its computational simplicity. This choice is rather 
inconsequential for our conclusions below, since we are 
introducing the $\left(\hat{s}^{1/2}\right)_{ave}$ variable only for
illustration purposes in Fig.~\ref{fig:ds}. As we shall see,
$\left(\hat{s}^{1/2}\right)_{ave}$ actually cancels out in the 
final comparison between the next two variables.
\item $\left(\hat{s}^{1/2}\right)_{thr}$ : this is the threshold 
of the true $\hat{s}^{1/2}$ distribution, i.e. the minimum 
allowed value of $\hat{s}^{1/2}$. Since the minimum 
$\hat{s}^{1/2}$ is obtained when the parent particles are 
produced at rest, $\left(\hat{s}^{1/2}\right)_{thr}$ is nothing but the 
sum of the parent particle masses, as indicated in 
eqs.~(\ref{sthr_gg}) and (\ref{sthr_cg}).
Therefore, $\left(\hat{s}^{1/2}\right)_{thr}$ is precisely the 
parameter that we would like to measure, 
in order to determine the true mass scale of the 
parent particles.
\item $\left(\hat{s}^{1/2}_{min}(2m_\chi)\right)_{peak}$ : 
this is the parameter defined in eq.~(\ref{speakdef}), 
namely the location of the peak of the $\hat{s}^{1/2}_{min}(2m_\chi)$
distribution, where we use the correct value for the 
invisible mass, in this case $M_{inv}=2m_\chi$, since
each SUSY event has two escaping LSPs.
\end{itemize}

\FIGURE[t]{
\epsfig{file=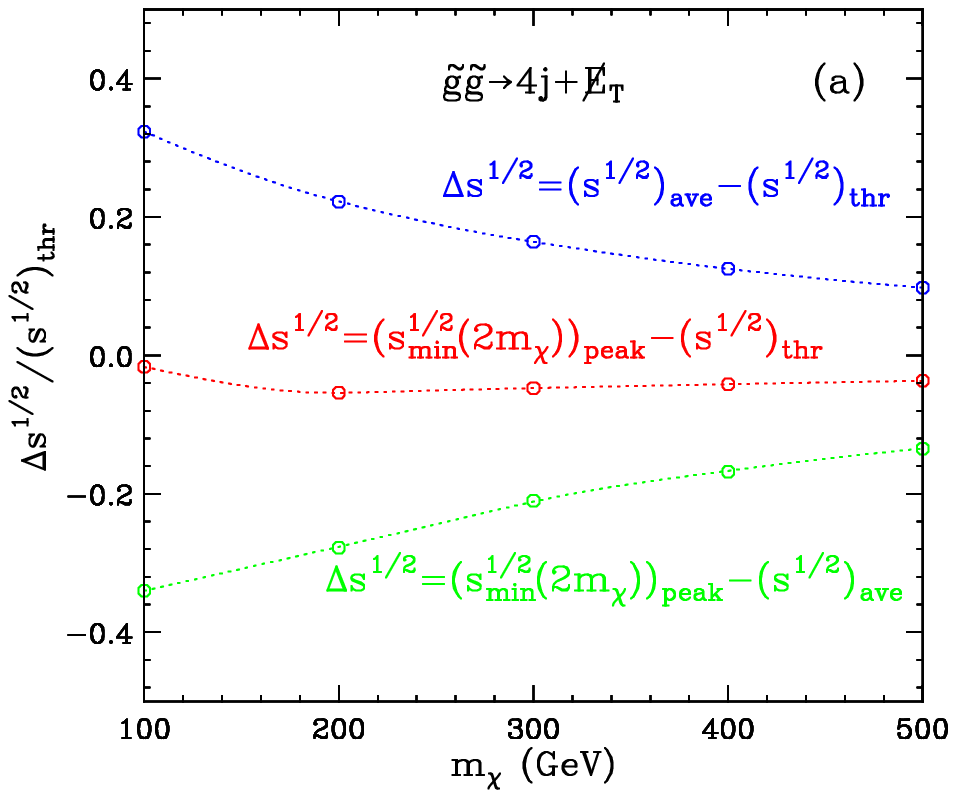,width=7.0cm}~~~
\epsfig{file=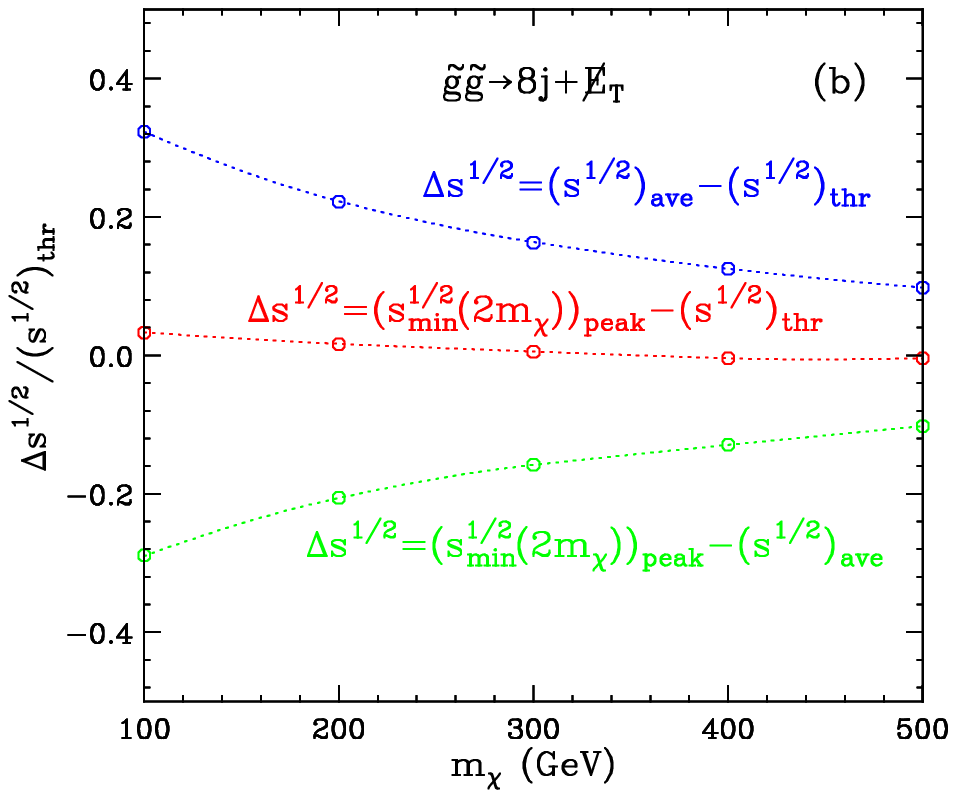,width=7.0cm}\\
\\
\epsfig{file=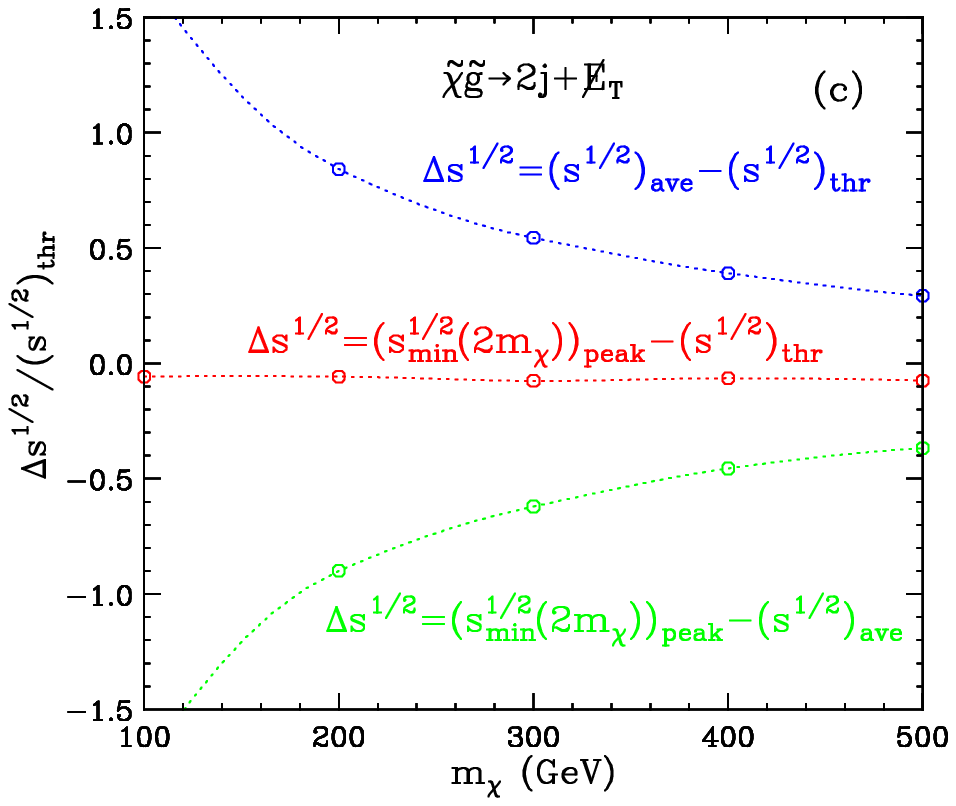,width=7.0cm}~~~
\epsfig{file=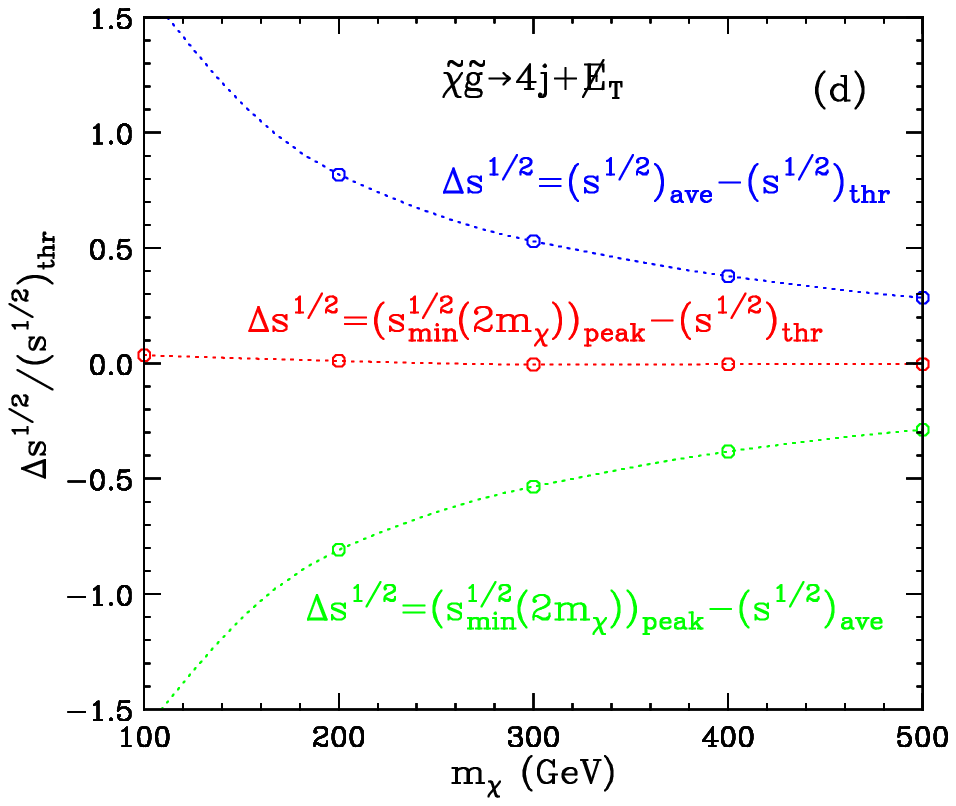,width=7.0cm}
\caption{\sl Validity of the approximation (\ref{sminsthr})
as a function of the LSP mass $m_\chi$. The
SUSY mass spectrum is fixed as in (\ref{gunif}).
In panels (a) and (b) we consider gluino pair production events,
while in panels (c) and (d) we study associated gluino-LSP production.
In panels (a) and (c) we force the gluino to decay to 2 jets as in
(\ref{2jet}), while in panels (b) and (d) each gluino decays to 4 jets as 
in (\ref{4jet}). In each panel we compare the following three quantities:
$\left(\hat{s}^{1/2}\right)_{ave}$, which is the average of the true $\hat{s}^{1/2}$ 
distribution; $\left(\hat{s}^{1/2}\right)_{thr}$, which is the threshold 
of the true $\hat{s}^{1/2}$ distribution; and 
$\left(\hat{s}^{1/2}_{min}(2m_\chi)\right)_{peak}$, 
which is the
location of the peak of the $\hat{s}^{1/2}_{min}(2m_\chi)$
distribution.
}
\label{fig:ds}}

According to our empirically derived conjecture (\ref{sminsthr}),
the last two variables are approximately equal, and the purpose of 
Fig.~\ref{fig:ds} is to test this hypothesis, using the previously
considered SUSY examples: gluino pair production (panels (a) and (b)),
and associated gluino-LSP production (panels (c) and (d)).
In panels (a) and (c) we force the gluino to decay to 2 jets as in
(\ref{2jet}), while in panels (b) and (d) each gluino decays to 4 jets as 
in (\ref{4jet}). Each line in Fig.~\ref{fig:ds} gives the fractional 
difference between a pair of $\hat{s}^{1/2}$ quantities as defined above.
For normalisation we used the value of $\left(\hat{s}^{1/2}\right)_{thr}$, 
which is given by (\ref{sthr_gg}) for panels (a) and (b) 
and by (\ref{sthr_cg}) for panels (c) and (d). 
We vary the relevant part of the SUSY spectrum by changing 
the input value of the LSP mass $m_\chi$ and adjusting the 
other masses in accord with (\ref{gunif}).

The main result in Fig.~\ref{fig:ds} is the comparison
between the experimentally observable quantity 
$\left(\hat{s}^{1/2}_{min}(2m_\chi)\right)_{peak}$
and the theoretical parameter $\left(\hat{s}^{1/2}\right)_{thr}$.
As indicated by the red lines in Fig.~\ref{fig:ds},
for the examples shown, those two quantities differ 
by no more than 10\%, thus validating our conjecture (\ref{sminsthr})
at the 10\% level as well. We find this result quite
intriguing. After all, we have not attempted any event 
reconstruction or decay chain identification,
we are looking at very complex and challenging multijet 
signatures, and we have even included detector resolution effects. 
After all those detrimental factors, the possibility of making 
any kind of statement regarding the mass scale of the 
new physics at the level of 10\% should be considered 
as rather impressive.

We find it instructive to understand how we ended up with
the observed precision, by comparing these two quantities 
$\left(\hat{s}^{1/2}_{min}(2m_\chi)\right)_{peak}$
and $\left(\hat{s}^{1/2}\right)_{thr}$ to the true $\hat{s}^{1/2}$
as represented by its average $\left(\hat{s}^{1/2}\right)_{ave}$.
The blue lines in Fig.~\ref{fig:ds} show the fractional 
difference between $\left(\hat{s}^{1/2}\right)_{ave}$ and $\left(\hat{s}^{1/2}\right)_{thr}$.
We see that this difference varies by quite a lot, on the order of
10-30\% for gluino pair-production, but may get in excess of
150\% for associated gluino-LSP production.
As expected, $\left(\hat{s}^{1/2}\right)_{ave}$ is always larger 
than the threshold value $\left(\hat{s}^{1/2}\right)_{thr}$, 
since the parent particles are typically produced
with some boost, and the blue lines in
Fig.~\ref{fig:ds} simply quantify the effect of this boost.

On the other hand, the green lines in Fig.~\ref{fig:ds}
represent the fractional difference (again 
normalised to $\left(\hat{s}^{1/2}\right)_{thr}$) between 
the measurable quantity 
$\left(\hat{s}^{1/2}_{min}(2m_\chi)\right)_{peak}$
introduced earlier in eq.~(\ref{speakdef}), and
the true energy scale of the events as given by
$\left(\hat{s}^{1/2}\right)_{ave}$. We see that this time the 
fractional difference is negative, which simply reflects 
the fact that our variable $\hat{s}^{1/2}_{min}$,
being defined through a minimization condition, 
will always underestimate the true energy scale. 
The interesting fact is that while the blue and 
green curves in Fig.~\ref{fig:ds} have opposite signs,
in absolute value they are very similar, leading to a 
fortuitous cancellation. The resulting discrepancy
indicated by the red lines is therefore much smaller 
than either of the two individual errors indicated by the
blue and green lines.

It is now easy to understand qualitatively the origin of 
the approximate relation (\ref{sminsthr}). Due to the 
boost at production, the true energy scale $\hat{s}^{1/2}$ 
is larger than the threshold energy $\left(\hat{s}^{1/2}\right)_{thr}$ 
by a certain amount. Later on, when we approximate $\hat{s}^{1/2}$
with $\hat{s}^{1/2}_{min}$, we underestimate the true energy scale 
$\hat{s}^{1/2}$ by more or less the same amount, 
bringing us back near the threshold $\left(\hat{s}^{1/2}\right)_{thr}$.
As a result, the $\hat{s}^{1/2}_{min}$ distribution
{\em peaks} very near the mass threshold $\left(\hat{s}^{1/2}\right)_{thr}$
which we are trying to measure in the first place.
Of course, the proximity of the $\hat{s}^{1/2}_{min}$ peak
to the threshold $\left(\hat{s}^{1/2}\right)_{thr}$ will be process
dependent, but according to the examples considered here,
holds to a remarkable accuracy. 

\section{Correlation of the $\hat{s}^{1/2}_{min}$ peak with the heavy particle mass threshold}
\label{sec:thr}

In the absence of a rigorous mathematical derivation,
eq.~(\ref{sminsthr}) should be considered simply as a conjecture.
Nevertheless, once eq.~(\ref{sminsthr}) is assumed to be 
approximately true, it allows us to {\em measure} 
the mass scale of the parent particles in terms of the 
hypothesized test mass $\tilde m_\chi$ of the lightest invisible particle,
e.g.~the LSP in SUSY. For example, in the case of gluino 
pair-production in SUSY, we can use eqs.~(\ref{sthr_gg}) 
and (\ref{sminsthr}) to obtain a measurement of the gluino mass
\beq
\tilde m_{\tilde g}(\tilde m_\chi) \approx 
\frac{1}{2} \left(\hat{s}^{1/2}_{min}(2\tilde m_\chi)\right)_{peak} 
\label{mgl_gg}
\eeq
as a function of the trial LSP mass $\tilde m_\chi$. Similarly, we 
can measure the gluino mass even in the 
much more challenging case of associated gluino-LSP production:
from eqs.~(\ref{sthr_cg}) and (\ref{sminsthr}), we obtain
\beq
\tilde m_{\tilde g} (\tilde m_\chi) \approx 
\left(\hat{s}^{1/2}_{min}(2\tilde m_\chi)\right)_{peak}
- \tilde m_\chi \ .
\label{mgl_cg}
\eeq
As evidenced from eqs.~(\ref{mgl_gg}) and (\ref{mgl_cg}),
these measurements are very straightforward, since 
the only experimental input needed for them is the
location of the peak of our all-inclusive global variable 
$\hat{s}^{1/2}_{min}$. One should not be bothered by 
the fact that we did not get an absolute measurement 
of the gluino mass, but only obtain it as a function of the 
LSP mass. This is a well-known drawback of the other
common mass measurement methods as well. For example,
the classic $M_{T2}$ endpoint analysis only yields the 
heavier parent mass as a function of the lighter child mass 
\cite{Lester:1999tx}. Similarly, the measurement of a single 
endpoint in some observable invariant mass distribution
provides only a single functional relation between the 
masses of the intermediate particles in the decay chain, 
and by itself does not measure the absolute scale. In this sense, 
our measurement (\ref{mgl_gg}) is on equal footing 
with the more traditional methods.

However, it is worth emphasizing the advantage
of our method in the case of asymmetric events, where
the parent particles are very different.
An extreme version of such events is provided by the
associated gluino-LSP production considered earlier.
Under those circumstances, the standard $M_{T2}$ method 
does not apply, while the single decay chain in the event 
may prove to be too short or too
messy to provide a clean measurement through the 
invariant mass endpoint method. In contrast, we can still
utilize $\hat{s}^{1/2}_{min}$ for the measurement 
indicated in (\ref{mgl_cg}) and a corresponding gluino mass 
determination.

\FIGURE[t]{
\epsfig{file=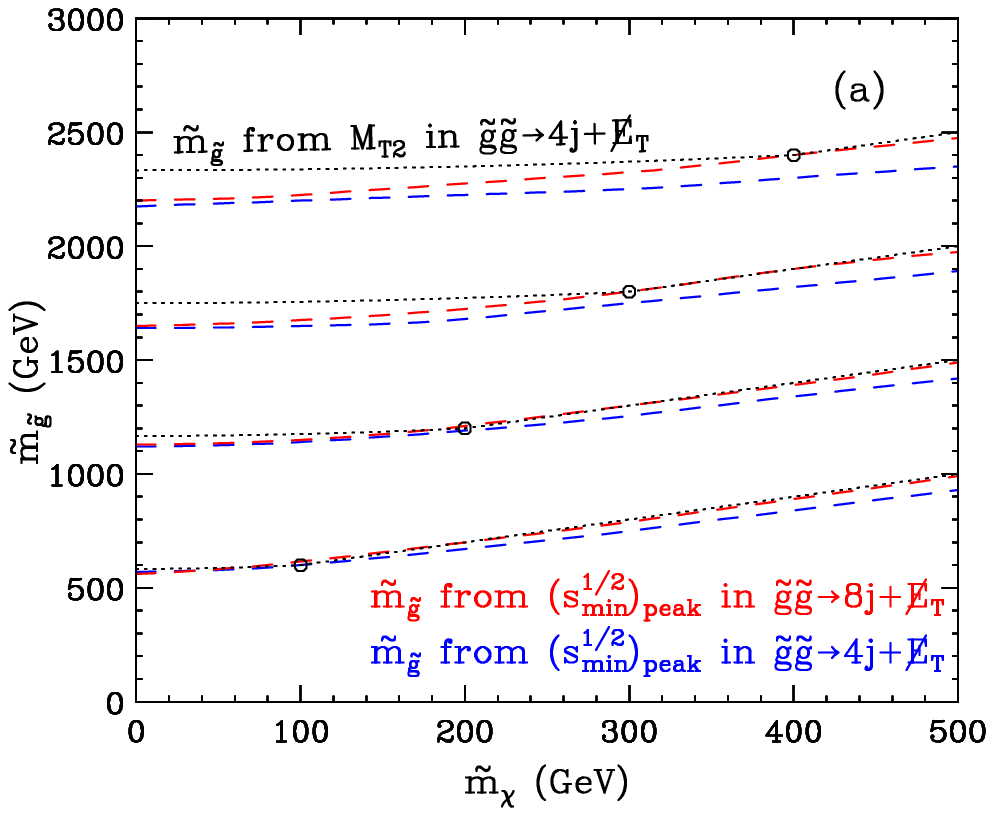,width=7.0cm}~~~
\epsfig{file=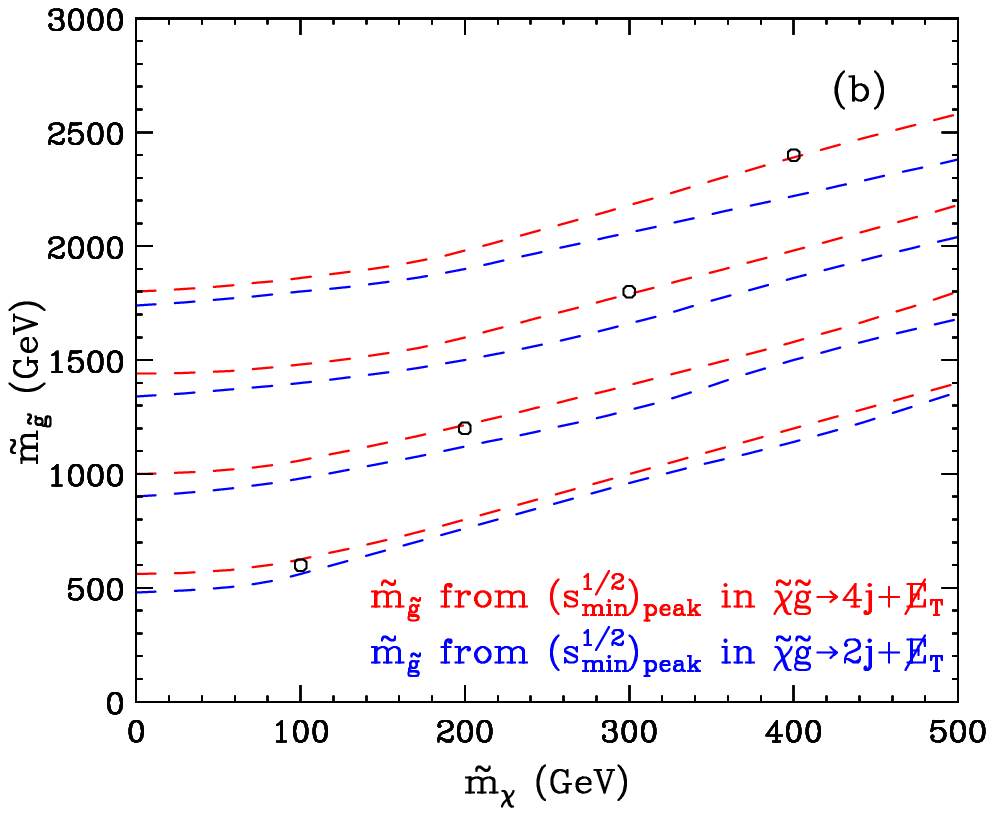,width=7.0cm}
\caption{\sl The correlation between the test
LSP mass $\tilde m_\chi$ and the corresponding
gluino mass $\tilde m_{\tilde g}$, derived from
(a) our proposed measurement (\ref{mgl_gg}) in gluino 
pair-production events, or 
(b) our proposed measurement (\ref{mgl_cg}) in associated gluino-LSP
production events.
Red (blue) lines correspond to the case of gluino decays 
to 4 jets as in (\ref{4jet})
(gluino decays to 2 jets as in (\ref{2jet})).
The black dotted lines in panel (a) indicate
the theoretically derived correlation 
from an ideal $M_{T2}$ endpoint analysis, 
i.e. assuming perfect resolution of the jet combinatorial ambiguity
and ignoring any detector smearing. 
The open circles mark the locations of
the true masses $(m_\chi,m_{\tilde g})$,
for each of our four study points.
}
\label{fig:meas}}

Let us now see how well the proposed
measurements (\ref{mgl_gg}) and (\ref{mgl_cg})
will do for each of the SUSY examples considered 
in the previous section. In Fig.~\ref{fig:meas}(a)
we used eq.~(\ref{mgl_gg}) to convert our previous 
measurements of the various 
$\hat{s}^{1/2}_{min}(2\tilde m_\chi)$ peaks 
in Figs.~\ref{fig:gg2j} and \ref{fig:gg4j} 
into a corresponding gluino mass measurement.
The red (blue) dashed lines correspond to
the case of 4-jet (2-jet) gluino decays as in
(\ref{4jet}) ((\ref{2jet})).
We show results for the same four study points 
used in the four panels of Figs.~\ref{fig:gg2j} 
and \ref{fig:gg4j}, and the open circles mark
the locations of the true masses $(m_\chi,m_{\tilde g})$,
for each study point. 

The quality of the measurement (\ref{mgl_gg})
can be judged from the proximity of the experimentally 
derived $\tilde m_{\tilde g}(\tilde m_\chi)$ curves
shown in the figure to the exact location of the 
true masses $(m_\chi,m_{\tilde g})$. We see that
both the red and blue curves in Fig.~\ref{fig:meas}(a)
pass very close to the true answer, especially for 
the study points with lower $m_\chi$.
In fact, %somewhat paradoxically, 
we obtain a better
measurement from the more complex 8-jet events
(the red curves). At first sight, 
this may seem counterintuitive, until one
realizes that the more visible objects are present in the event, 
the smaller the effect of the missing particles, 
and hence the smaller the error due to our approximation (\ref{ptsol}, \ref{pzsol}).
Such multijet events appear very challenging 
to be tackled by any other means. 
For the sake of comparison, the black dotted lines
in Fig.~\ref{fig:meas}(a) show
the theoretically derived correlation 
from an ideal $M_{T2}$ endpoint analysis, 
i.e. assuming perfect resolution of the jet combinatorial ambiguity
and ignoring any detector resolution effects. 
Comparing the red line from our measurement (\ref{mgl_gg}) 
to the ideal $M_{T2}$ line, we are tempted to conclude that,
in essence, our $\hat{s}^{1/2}_{min}$ variable contains
pretty much the same amount of information as $M_{T2}$.
The big advantage of $\hat{s}^{1/2}_{min}$, however, is the
fact that we can obtain this information at a much lower 
cost in terms of analysis effort.

Finally, in Fig.~\ref{fig:meas}(b) we show our results
from the analogous measurement (\ref{mgl_cg}) in the case of
associated gluino-LSP production. Here we also consider two
different options for the gluino decay ---
2 jet decays as in (\ref{2jet}) (blue lines), or
4 jet decays as in (\ref{4jet}) (red lines). 
We then plot the resulting functional dependence 
$\tilde m_{\tilde g}(\tilde m_\chi)$ for each of the 
four study points considered earlier.
Comparing Fig.~\ref{fig:meas}(b) to Fig.~\ref{fig:meas}(a)
which we just discussed, we arrive at very similar conclusions:
the measurement (\ref{mgl_cg}) is still quite accurate,
and the superior result is provided by the more complex 
topology. Notice that here we do not show any $M_{T2}$-based
results, since the concept of $M_{T2}$ can not be applied to 
an extremely asymmetric topology like this one.

\section{The impact of initial state radiation and multiple parton interactions}
\label{sec:ISR}

Up to now we have been discussing the $\sqrt{\hat s}$ variable 
of the primary parton-level hard scattering (HS).
In principle, $\sqrt{\hat s}$ can be measured exactly, 
whenever we could both detect and identify the decay products 
of the heavy particles which were initially produced in the HS.
Unfortunately, in reality it is rather difficult to measure
$\sqrt{\hat s}$ directly, for a couple of reasons:
\begin{enumerate}
\item {\em Omitting relevant particles from the $\sqrt{\hat s}$ calculation.}
This case arises whenever some of the decay products resulting from 
the HS are not detected. For example, this may happen due to the 
imperfect hermeticity of the detector, where some of the relevant 
decay products are lost down the beam pipe. Fortunately, in reality 
this effect is pretty small. A much more serious problem arises whenever there are
invisible particles $\chi_i$ (see Fig.~\ref{fig:metevent})
among the relevant decay products. Then, a relatively large fraction 
of the initial $\sqrt{\hat s}$ may go undetected, as can be seen by
comparing the $\sqrt{\hat s}_{min}$ distributions in Figs.~\ref{fig:varsgg}-\ref{fig:cg4j}
to the respective true (yellow-shaded) $\sqrt{\hat s}$ distributions.
\item {\em Including irrelevant particles in the $\sqrt{\hat s}$ calculation.}
In general, any given event will contain a certain number of particles 
which will be seen in the detector, but {\em did not} originate from
the primary HS. Initial state radiation (ISR), multiple parton interactions
(MPI) and pile-up are the main examples of processes contributing to this effect.
The pile-up effect can be controlled by a suitable $\Delta z$ cut, removing
from consideration tracks which do not appear to originate from the 
primary vertex. However, ISR and MPI can be a serious problem.
Including the extra particles will necessarily lead to an {\em increase}
in the measured value of $\sqrt{\hat s}$. In order to emphasize this
difference, in the rest of this section we shall be using a prime 
to designate the experimentally measured quantities which include 
the full ISR and MPI effects 
($\sqrt{{\hat s}'}$ and $\sqrt{{\hat s}'}_{min}$, correspondingly).
\end{enumerate}
Our proposal for dealing with the first of these two problems 
was to introduce the $\sqrt{\hat s}_{min}$ variable in lieu of
the true $\sqrt{\hat s}$. We then found an interesting empirical 
correlation (\ref{sminsthr}) between $\sqrt{\hat s}_{min}(2m_\chi)$
and the new physics mass scale. Now we shall turn our attention to 
dealing with the second problem, namely the fact that 
$\sqrt{{\hat s}'}_{min} > \sqrt{\hat s}_{min}$.

Before we begin, we should mention that, depending on the particular 
circumstances and/or the goal of the experimenter, there may be certain 
situations where the inequality $\sqrt{{\hat s}'}_{min} > \sqrt{\hat s}_{min}$
may not represent an actual problem. For example, if one 
is simply trying to measure the {\em total} energy  
in the observed events and not just the energy of the HS,
then for missing energy events the relevant quantity of 
interest would be $\sqrt{{\hat s}'}_{min}$ itself, which
would still be given by the expression (\ref{smin_def})
derived in Section \ref{sec:derivation}. There may also be
situations where the ISR and/or MPI products may be reliably 
identified and excluded from the $\sqrt{\hat s}_{min}$
calculation. For example, consider a lepton collider and
a missing energy signature with any number of jets
and/or leptons. Since MPI is absent, while ISR and beamstrahlung 
would only contribute photons, there will be no confusion with regards 
to which particles are due to ISR and which are coming from the HS. 
The analogous example at hadron colliders would be a signature
containing anything but QCD jets. In what follows we shall 
ignore such trivial cases and instead focus on the much more
challenging case of hadron colliders and jetty signatures, 
where the ISR/MPI products cannot be easily recognized.

In the absence of any reliable methods for resolving the jet combinatorial 
problem on an event by event basis, one is left with two options.
First, one may try to compensate for the ISR/MPI effects 
on the global $\sqrt{{\hat s}'}_{min}$ distribution.
In order to do this, one needs to know how ISR/MPI would
affect the original $\sqrt{\hat s}_{min}$ distribution. Ideally,
this information should be measured from real data, using some
Standard Model process as a standard candle.
For example, Drell-Yan can provide the relevant 
information for a $q\bar{q}$ initial state \cite{Kar:2008zz},
while $t\bar{t}$ can be used to study the $gg$ initial state.
Alternatively, one may calculate the ISR effects from first 
principles in QCD. Both of these approaches will be 
pursued in a future work \cite{KKMPW}.

A second approach would be to design and apply cuts which would
minimize the ISR and MPI effects on the calculation of $\sqrt{{\hat s}'}_{min}$.  
Unfortunately, this is rather difficult to do in a 
model-independent fashion, since the size 
of the ISR effect is very model-dependent and depends 
on many factors: the energy of the collider (Tevatron  or LHC),
the mass of the produced particles,  the identity of the 
partons initiating the HS, etc. Therefore, the optimal 
method to compensate for the ISR effect will also depend on all 
of these factors and will need to be decided on a case by case
basis.

\FIGURE[t]{
\epsfig{file=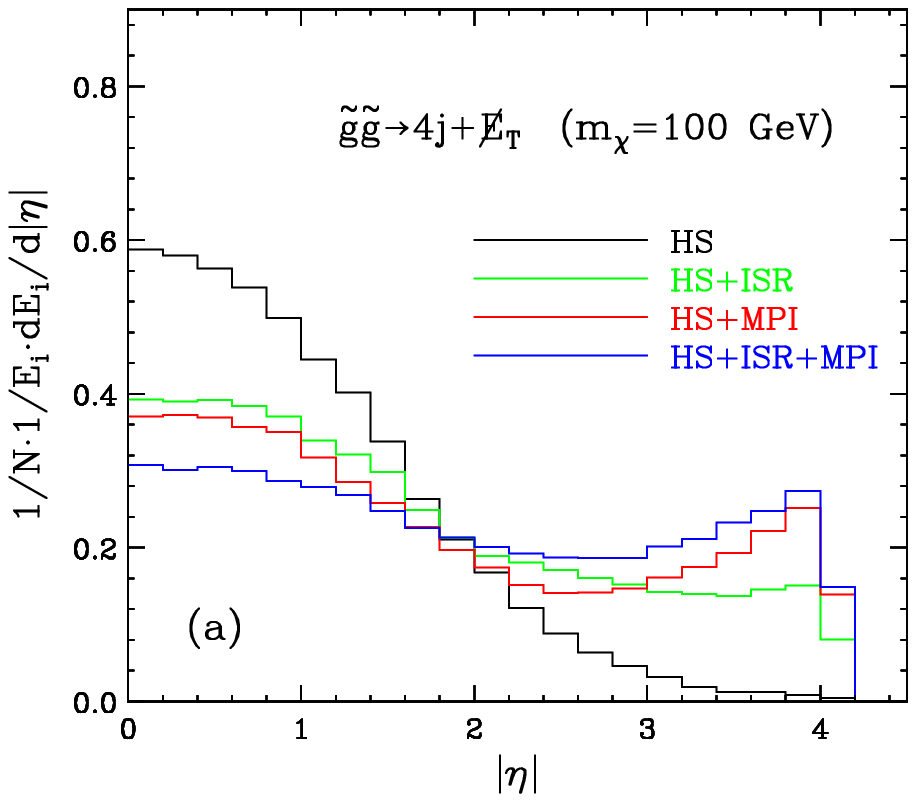,width=7.0cm}~~~
\epsfig{file=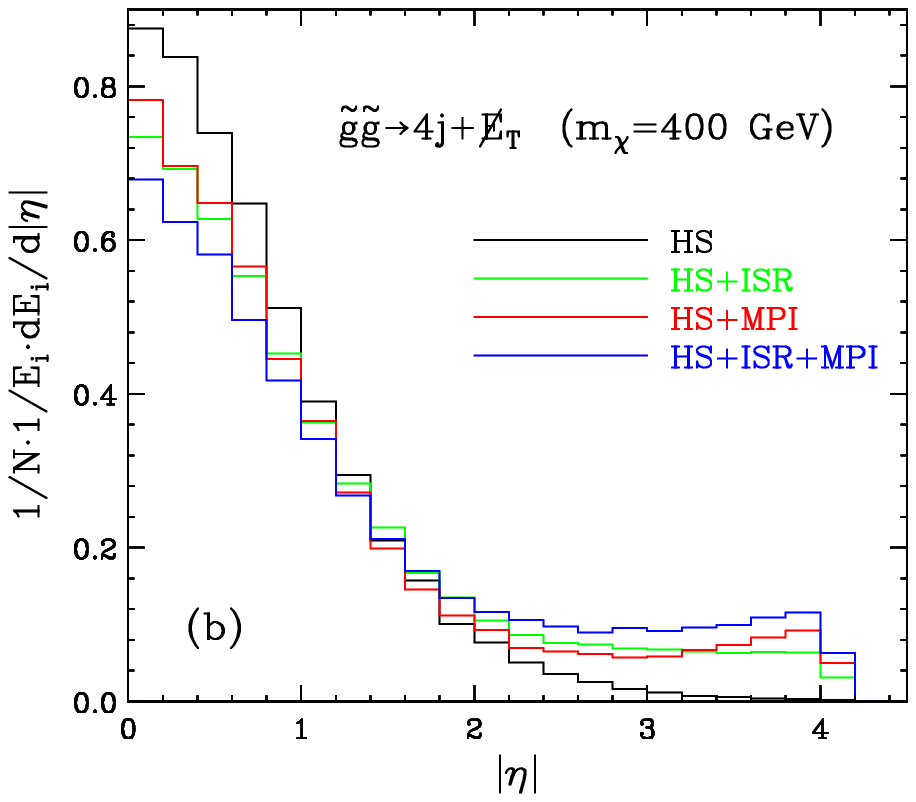,width=7.0cm}\\
\\
\epsfig{file=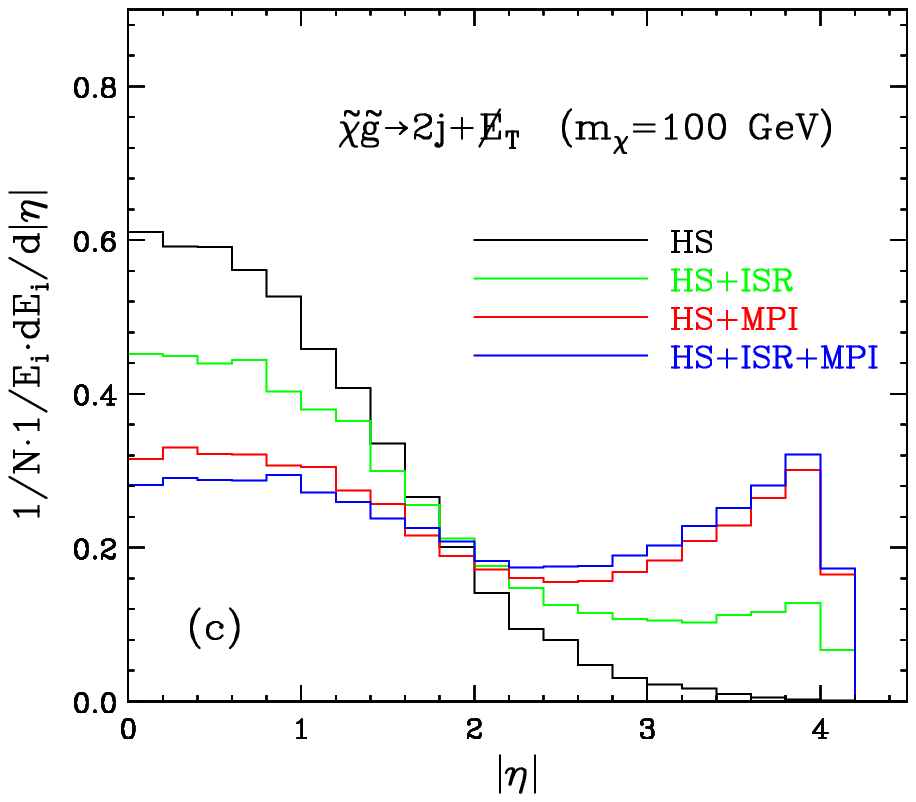,width=7.0cm}~~~
\epsfig{file=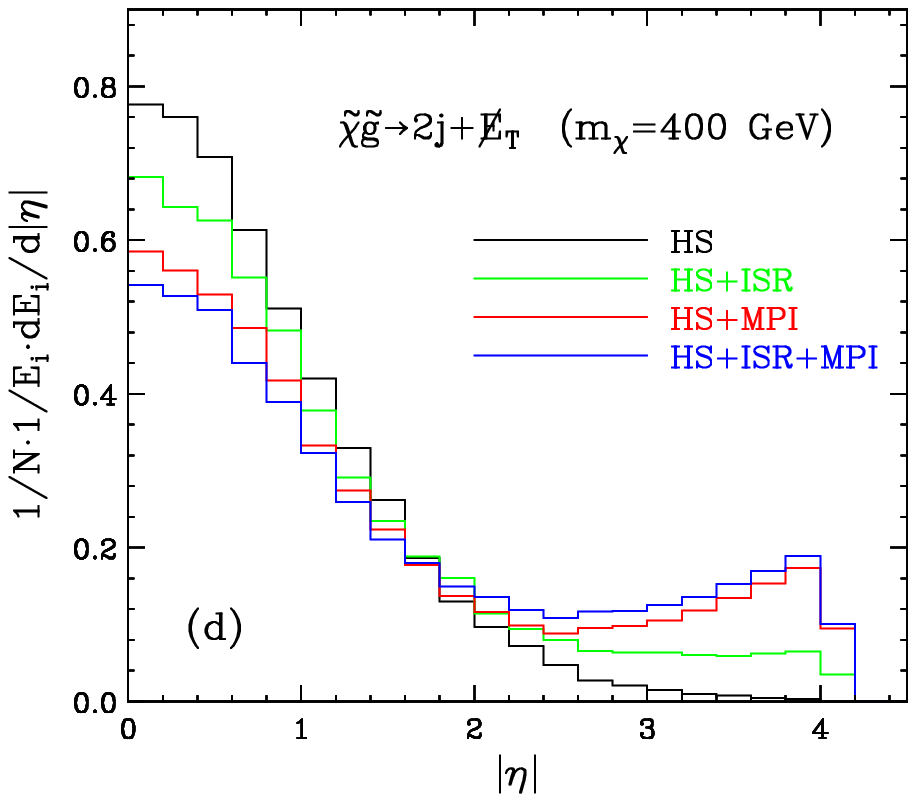,width=7.0cm}
\caption{\sl Energy distributions as a function of $|\eta|$,
for some of the SUSY examples considered earlier: 2-jet gluino decays 
from gluino pair production with 
(a) $m_\chi=100$ GeV or (b) $m_\chi=400$ GeV; 
and from associated gluino-LSP production with 
(c) $m_\chi=100$ GeV  or (d) $m_\chi=400$ GeV.
The color scheme is such that the black histograms 
correspond to our previous results from Section \ref{sec:mass} 
in the idealised case without ISR or MPI,
while the green (red, blue) histograms 
include the effect of ISR (MPI, both ISR and MPI).
Here $E_i$ is the total energy measured in the $i$-th event, and 
$N$ is the total number of events. As a result, all distributions 
shown in the figure are unit-normalized.
}
\label{fig:dE_deta}}
%
%
%
%Nevertheless, at this point it is worth asking whether
%something can be done to reduce the ISR and MPI effects seen in 
%Fig.~\ref{fig:dN_dsp}, which are quite significant. 
For the purposes of the current study, we shall use
a simple cut-based approach as discussed here, postponing 
the more complete treatment for \cite{KKMPW}.
To this end, we need to identify some global 
property of the ISR and MPI products which would distinguish them from the HS.
Since it is well known that ISR and MPI peak in the forward region, 
it is natural to consider the pseudorapidity $\eta$ as a 
simple cut variable. The energy distributions as a function of $|\eta|$,
for a few representative cases are shown in Fig.~\ref{fig:dE_deta}.
We again consider the processes of gluino pair production 
(Figs.~\ref{fig:dE_deta}(a) and \ref{fig:dE_deta}(b))
and associated gluino-LSP production 
(Figs.~\ref{fig:dE_deta}(c) and \ref{fig:dE_deta}(d)).
In each case, the gluino decays to 2 jets as in (\ref{2jet}).
We choose to show the two extreme cases for the mass spectrum
considered earlier: 
$m_\chi=100$ GeV (Figs.~\ref{fig:dE_deta}(a) and \ref{fig:dE_deta}(c)) and
$m_\chi=400$ GeV (Figs.~\ref{fig:dE_deta}(b) and \ref{fig:dE_deta}(d)).
The gluino mass is still fixed according to the gaugino 
unification relation (\ref{gunif}).
The black histograms in Fig.~\ref{fig:dE_deta} represent
our previous results from Section \ref{sec:mass}
without any ISR or MPI effects,
while the green (red) histograms include the effect of ISR (MPI).
Finally, the blue histograms include both the ISR and MPI effects.
The plots in Fig.~\ref{fig:dE_deta} are normalized as follows. 
For each event, say the $i$-th one, we add the energy deposits in all 
calorimeter towers at a given $|\eta|$, then divide the sum by the 
total energy $E_i$ observed in the $i$-th event and the 
total number of events $N$, and finally enter the result
into the corresponding $|\eta|$ bin. It is easy to see that 
this ensures that the final distributions are unit-normalized.

Fig.~\ref{fig:dE_deta} shows that, as expected, the ISR and MPI effects
appear mostly in the forward region. Therefore, by applying a simple 
$|\eta|<\eta_{max}$ cut, we could reduce their impact.
Of course any such rapidity cut would essentially bring us back
closer to the transverse quantities from which we were trying to escape
from the very beginning. Furthermore, such a simple-minded procedure would
introduce an uncontrollable systematic error, which would have to be
estimated on a case by case basis. For example, Fig.~\ref{fig:dE_deta}(b)
shows that when the spectrum is rather heavy,
the ISR/MPI effects are relatively small and can probably be safely 
neglected altogether, while Figs.~\ref{fig:dE_deta}(a), 
\ref{fig:dE_deta}(c) and \ref{fig:dE_deta}(d)
reveal a significant ISR/MPI pollution for a light SUSY spectrum. 
One should also keep
in mind that our conjecture (\ref{sminsthr}) is already subject to
a certain systematic error, whose size sets the benchmark for the 
ISR/MPI elimination study. With those caveats, we choose our cut 
at $\eta_{max}=1.4$, which is nothing but the end of the
barrel and beginning of HE/HF calorimeters in CMS.
This choice makes good sense from an experimentalist's point of view,
since the segmentation and performance of the HE/HF calorimeters are
relatively worse to begin with.

\FIGURE[t]{
\epsfig{file=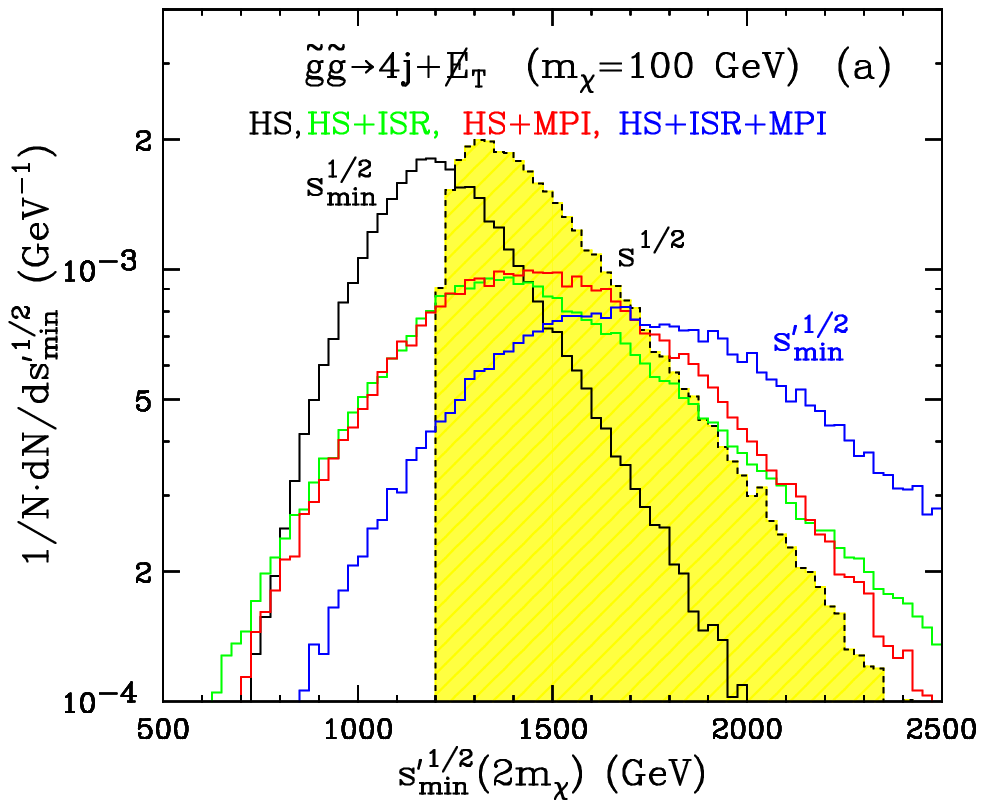,width=7.0cm}~~~
\epsfig{file=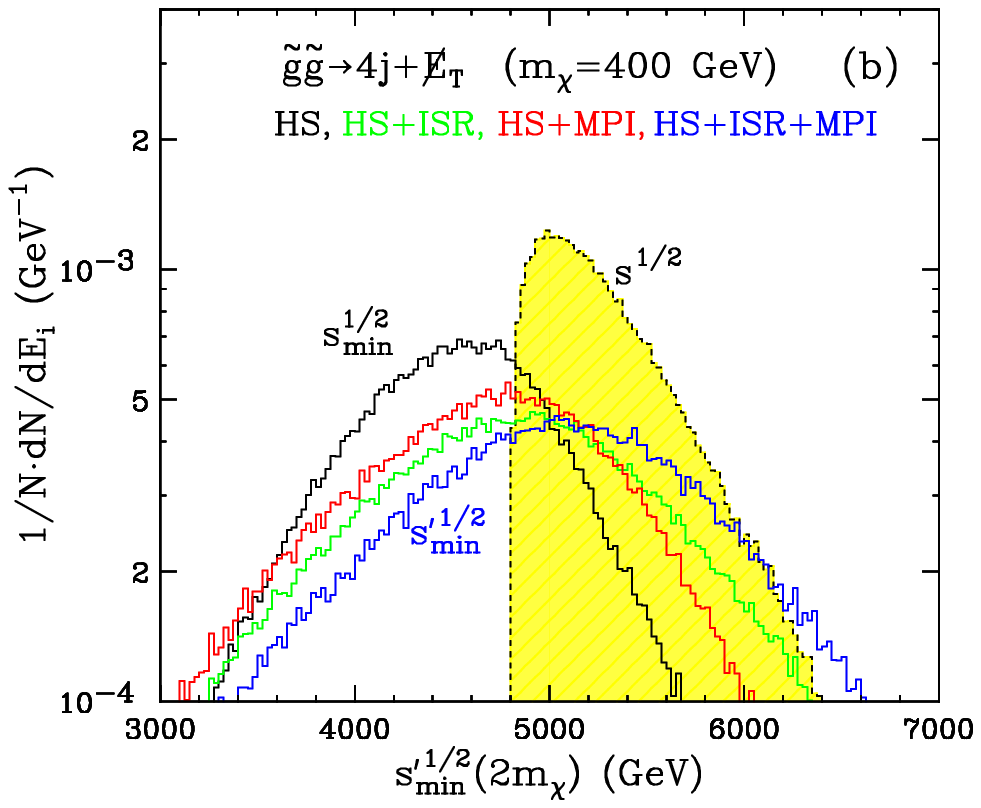,width=7.0cm}\\
\\
\epsfig{file=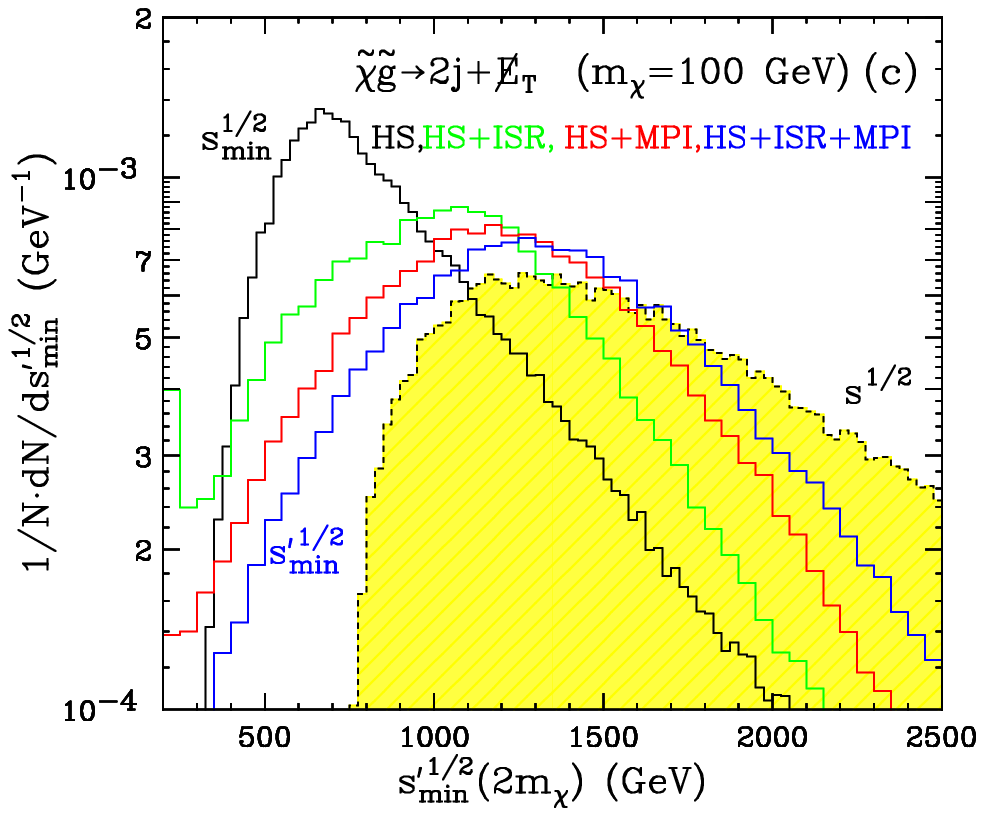,width=7.0cm}~~~
\epsfig{file=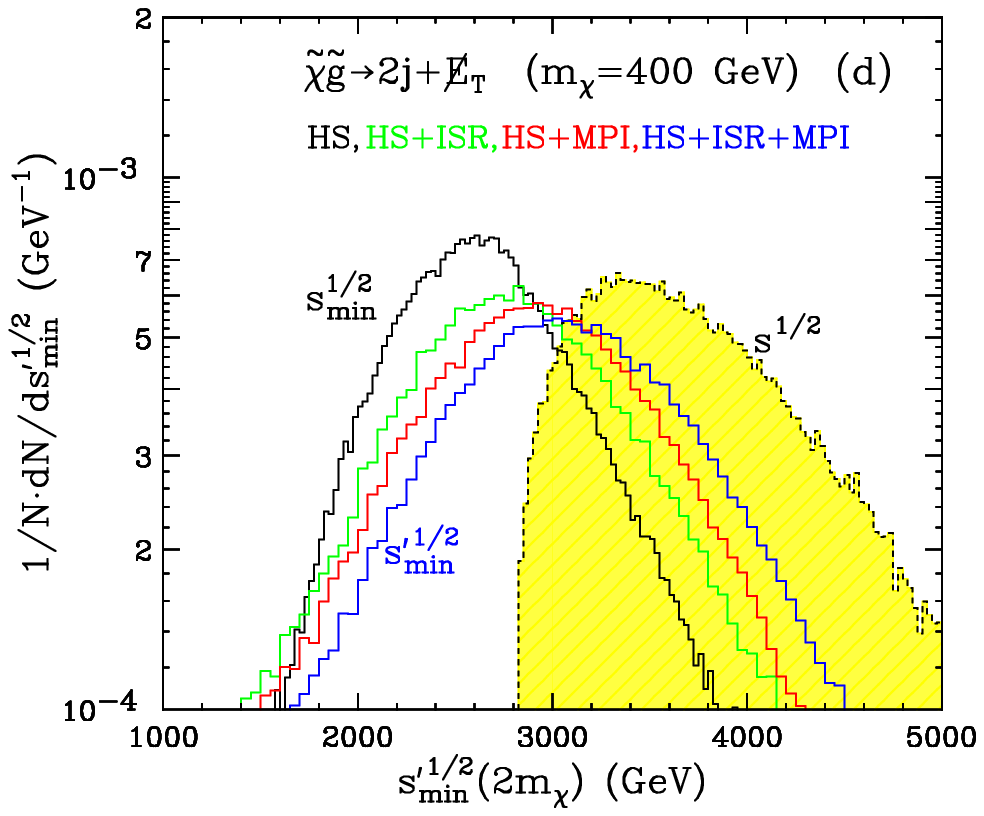,width=7.0cm}
\caption{\sl Unit-normalized distributions of
$\sqrt{{\hat s}}_{min}(2m_\chi)$ and $\sqrt{{\hat s}'}_{min}(2m_\chi)$
for the SUSY examples considered in Fig.~\ref{fig:dE_deta}. 
The color scheme is the same as in Fig.~\ref{fig:dE_deta}. 
%for some of the SUSY  
%examples considered earlier: 2-jet gluino decays 
%from gluino pair production with (a) $m_\chi=100$ GeV or 
%(b) $m_\chi=400$ GeV; 
%and from associated gluino-LSP production with 
%(c) $m_\chi=100$ GeV  or (d) $m_\chi=400$ GeV.
%The color scheme is such that the black histogram is
%our previous result for $\sqrt{{\hat s}}_{min}(2m_\chi)$
%from Section \ref{sec:mass} in the idealised case 
%without ISR or MPI, while the green (red) histogram 
%includes the effect of ISR (MPI). The blue histogram
%includes both the ISR and MPI effects, and
%represents the actually measured $\sqrt{\hat s'}_{min}(2m_\chi)$.
The blue histograms include both the ISR and MPI effects, and
represent the actually measured $\sqrt{\hat s'}_{min}(2m_\chi)$,
while the green (red) histograms include the effects of ISR (MPI)
only. All three of those distributions are subject to the 
$|\eta|<1.4$ cut discussed in the text. 
For comparison, we also show our previous results
from Section \ref{sec:mass}, corresponding to the HS only
(without any ISR or MPI effects) and
without an $\eta$ cut. In particular, the
black solid histograms in Fig.~\ref{fig:dN_dsp} represent 
our previous results for the quantity $\sqrt{\hat s}_{min}(2m_\chi)$, while
the black dotted (yellow-shaded) histograms give the true 
$\hat{s}^{1/2}$ distribution, whose threshold is the 
parameter to be measured.
}
\label{fig:dN_dsp}}

Let us now revisit some of the $\sqrt{\hat s}_{min}$ distributions
from Section~\ref{sec:mass} and incorporate successively the 
effects of ISR and/or MPI. 
Fig.~\ref{fig:dN_dsp} shows our results for the same four SUSY examples from
Fig.~\ref{fig:dE_deta}.
%Results for a few representative cases
%are contained in Fig.~\ref{fig:dN_dsp}, which shows unit-normalized 
%distributions of $\sqrt{{\hat s}'}_{min}(2m_\chi)$.
%We again consider the processes of gluino pair production 
%(Figs.~\ref{fig:dN_dsp}(a) and \ref{fig:dN_dsp}(b))
%and associated gluino-LSP production 
%(Figs.~\ref{fig:dN_dsp}(c) and \ref{fig:dN_dsp}(d)).
%In each case, the gluino decays to 2 jets as in (\ref{2jet}).
%We choose to show the two extreme cases for the mass spectrum
%considered earlier: 
%$m_\chi=100$ GeV (Figs.~\ref{fig:dN_dsp}(a) and \ref{fig:dN_dsp}(c)) and
%$m_\chi=400$ GeV (Figs.~\ref{fig:dN_dsp}(b) and \ref{fig:dN_dsp}(d)).
%The gluino mass is still fixed according to the gaugino 
%unification relation (\ref{gunif}).
The green (red) histograms include the effect of ISR (MPI) alone, 
while the blue histograms include both the ISR and MPI effects, 
and thus represent the true measured quantity $\sqrt{\hat s'}_{min}(2m_\chi)$.
All three of those distributions are subject to our $|\eta|<1.4$ cut. 
For comparison, we also show our previous results
from Section \ref{sec:mass}, corresponding to the HS only
(without any ISR or MPI effects) and
without an $\eta$ cut. In particular, the
black solid histograms in Fig.~\ref{fig:dN_dsp} represent 
our previous results for the quantity $\sqrt{\hat s}_{min}(2m_\chi)$, while
the black dotted (yellow-shaded) histograms give the true 
$\hat{s}^{1/2}$ distribution, whose threshold is the 
parameter that ideally we would like to measure.

Fig.~\ref{fig:dN_dsp} confirms that the ISR and MPI effects 
shift the original HS distribution $\sqrt{\hat s}_{min}$ (black histograms)
into a harder $\sqrt{\hat s'}_{min}$ distribution (blue histograms),
even after applying the $\eta$ cut.
The size of this effect depends on the mass spectrum: 
it is more pronounced 
when the spectrum is light\footnote{Notice that for a 
given value of $m_\chi$, the relevant mass scale 
$2m_{\tilde g}=12m_\chi$ in Figs.~\ref{fig:dN_dsp}(a) and \ref{fig:dN_dsp}(b)
is almost twice as large as the corresponding mass scale
$m_{\tilde g}+m_{\tilde\chi^0_1}=7m_\chi$  in Figs.~\ref{fig:dN_dsp}(c) and \ref{fig:dN_dsp}(d).
}, as in Figs.~\ref{fig:dN_dsp}(a),
\ref{fig:dN_dsp}(c) and \ref{fig:dN_dsp}(d).
%and when the signature is relatively simple ($2j+\met$ as opposed to
%$4j+\met$). 
In the worst case scenario of Fig.~\ref{fig:dN_dsp}(c)
the location of the $s^{1/2}_{min}$ 
peak shifts by almost a factor of two. On the other hand, for the
best-case scenario of Fig.~\ref{fig:dN_dsp}(b) the shift is rather 
small. By comparing the green and red histograms, we can also 
deduce the relative importance of ISR versus MPI. We see that
the two effects are roughly comparable in size, but
as a rule, the red histograms are shifted further along, which 
suggests that MPI has a somewhat higher impact than ISR, indicating
the importance of understanding the full structure of the 
underlying event at the LHC. The general conclusion from
Fig.~\ref{fig:dN_dsp} is that our mass measurement method 
proposed in Section~\ref{sec:mass} is likely to work much better 
if the new particle spectrum happens to be relatively heavy. 
This assumption is not unreasonable: 
%recall that our main motivation
%for introducing $\sqrt{\hat s}_{min}$ was precisely the case when
%the signature is too complex to tackle by exclusive methods.
%Furthermore, 
if the new physics spectrum were too light, then 
it might have already been ruled out directly or indirectly, 
and if not, then due to the higher production cross-sections,
there should at least be sufficient 
statistics to attempt some sort of exclusive reconstruction.
In this sense, for the case where $\sqrt{\hat s}_{min}$ is most 
likely to be useful, ISR and MPI are least likely to be a problem.

\FIGURE[t]{
\epsfig{file=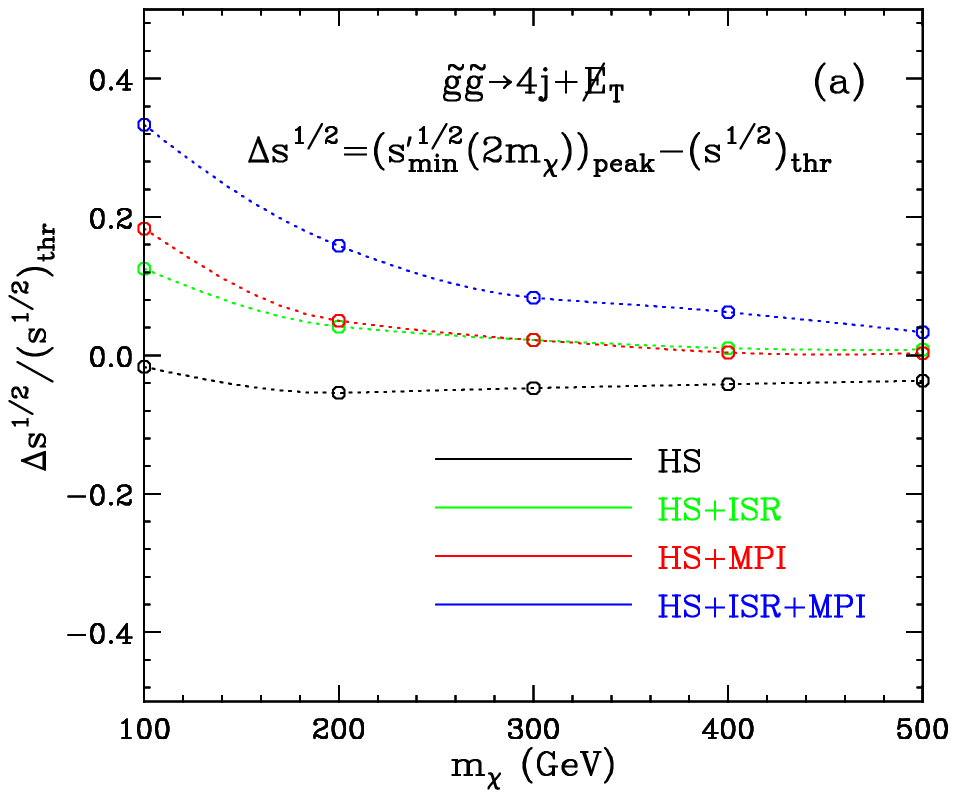,width=7.0cm}~~~
\epsfig{file=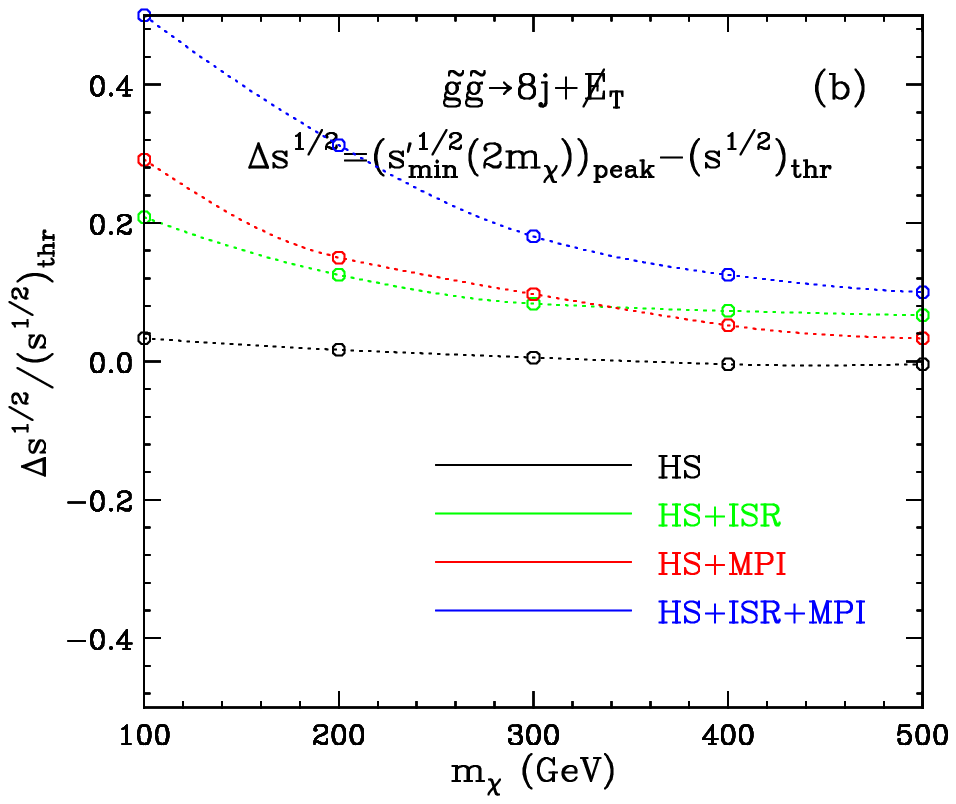,width=7.0cm}\\
\\
\epsfig{file=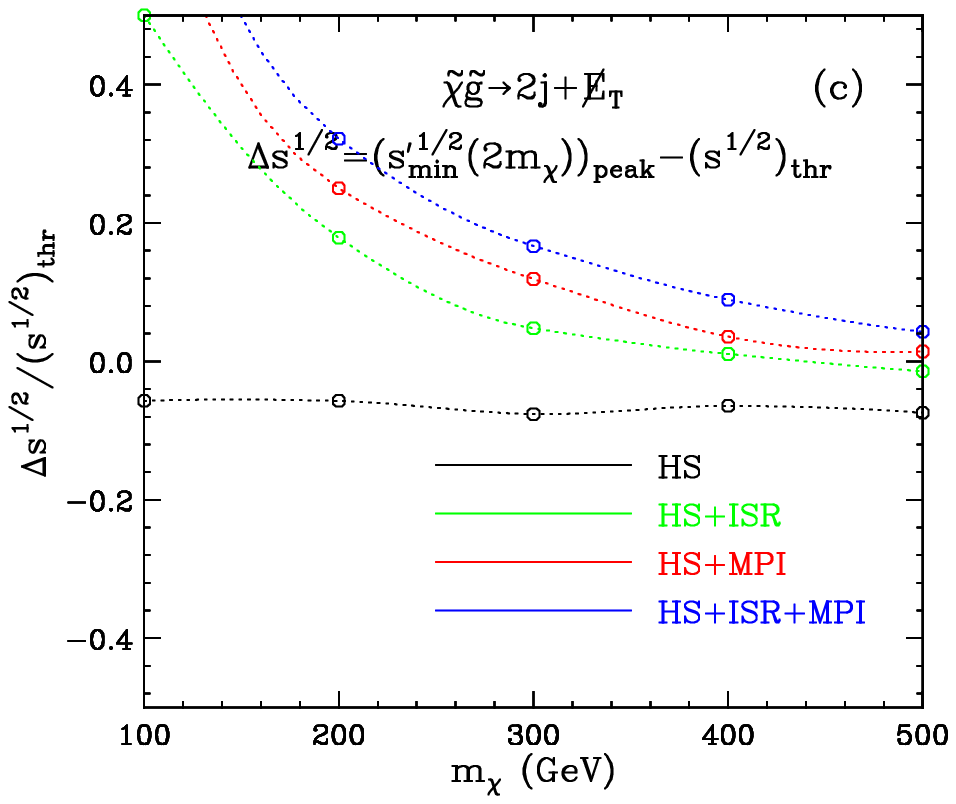,width=7.0cm}~~~
\epsfig{file=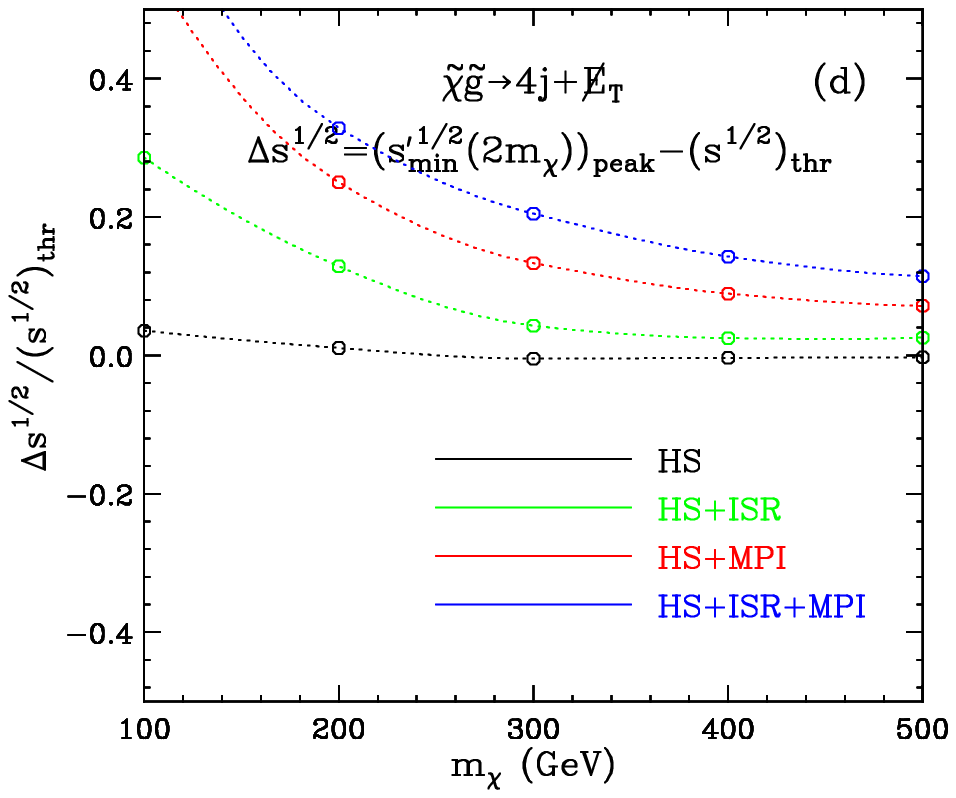,width=7.0cm}
\caption{\sl The same as Fig.~\ref{fig:ds}, but including
the effects of ISR (green), MPI (red), both ISR and MPI (blue).
The variable ${\hat{s}'{}}^{1/2}_{\!\! min}(2m_\chi)$ here is calculated 
with a cut of $|\eta|<1.4$, corresponding to the coverage of
the CMS barrel calorimeter only.}
\label{fig:ds_ISR}}

We are now in a position to repeat our mass measurement analysis 
from Section~\ref{sec:thr}, with the inclusion of ISR and MPI, 
while ignoring the forward calorimetry through an $|\eta|<1.4$ cut.
Our results are shown in Fig.~\ref{fig:ds_ISR}.
Comparing Figs.~\ref{fig:ds} and \ref{fig:ds_ISR}, we see that 
the inclusion of ISR/MPI deteriorates the mass measurement, 
most notably for light SUSY mass spectra with $m_\chi\sim 100-200$ GeV.
This should not be surprising, given what we have already seen
in Figs.~\ref{fig:dE_deta} and \ref{fig:dN_dsp}.
Nevertheless, for heavier SUSY spectra the precision remains 
relatively good, typically on the order of $10\%$, even for the 
most challenging cases of associated gluino-LSP production.

\section{Summary and conclusions}
\label{sec:summary}

Anticipating that an early (late) discovery of a missing energy 
signal at the LHC (Tevatron) may involve a signal topology which 
is too complex for a successful and immediate exclusive event 
reconstruction, we proposed a new global and inclusive 
variable $\hat{s}^{1/2}_{min}$, defined as follows: it is 
the minimum required center-of-mass energy, given the measured 
values of the total calorimeter energy E, total visible momentum
$\vec{P}$, and/or missing transverse energy $\met$ in the event.
Our variable has several desirable features:
\begin{itemize}
\item It is {\em global} in the sense that it uses all of the 
available information in the event and not just transverse quantities,
for example. 
\item It is {\em inclusive} in the sense that it does not depend 
on the specific production process, or particular decay chain.
Consequently, it is also very model-independent and does not
require any exclusive event reconstruction, which may be 
a great advantage in the early days of the LHC.
\item It is theoretically well defined and as such has a clear 
physical meaning: it gives the minimum total energy which is 
consistent with a given observed event. This intuitively 
clear physical picture allowed us to correlate it with 
the mass threshold of the new particles as in eq.~(\ref{sminsthr}),
which turned out to work surprisingly well. In contrast,
it is generally difficult to correlate a bump in a 
purely transverse quantity like $\met$ or $H_T$ to 
any physical mass parameter in a model-independent fashion.
\end{itemize}

In Section~\ref{sec:derivation} we derived a simple formula (\ref{smin_def})
for $\hat{s}^{1/2}_{min}$ in terms of the measured $E$, $P_z$ and $\met$. 
The formula is in fact completely general, and is valid 
for any generic event shown in Fig.~\ref{fig:metevent},
with an arbitrary number and/or types of missing particles.
Therefore, it can be applied equally successfully to SM 
as well as BSM missing energy signals. 

In Sections \ref{sec:compare} and \ref{sec:mass} we
identified two useful properties of the $\hat{s}^{1/2}_{min}$
variable. First, its shape matches the true $\hat{s}^{1/2}$
distribution better than any of the other global 
inclusive quantities which are commonly discussed in the literature.
More importantly, when we create the $\hat{s}^{1/2}(M_{inv})$
distribution with the true value of the invisible mass $M_{inv}$,
its peak is very close to the mass threshold of the 
parent particles originally produced in the event.
This conjecture, summarized in eq.~(\ref{sminsthr}),
allows us to obtain a rough estimate of the 
new physics mass scale, as a function of the single parameter 
$M_{inv}$. For example, in $R$-parity conserving supersymmetry, 
where $M_{inv}=2m_\chi$, we derive a relation between 
the heavy superpartner mass and the mass of the LSP,
as shown in Fig.~\ref{fig:meas}.

Before we conclude, we should comment on several other potential 
uses of the $\hat{s}^{1/2}_{min}$ variable. Before we even get to
the discovery stage, $\hat{s}^{1/2}_{min}(0)$ can already be 
used for background rejection and increasing signal to noise,
just like $M_{T2}(0)$ \cite{Barr:2005dz}. In particular, it
is interesting to explore the {\em correlations} between 
$\hat{s}^{1/2}_{min}$ and the other global inclusive 
variables discussed in Section \ref{sec:compare} \cite{Alwall:2008va}.
While we did not include any SM backgrounds in our SUSY plots, 
we expect that
the presence of SM backgrounds will not affect either the 
existence or the location of the new physics $\hat{s}^{1/2}_{min}(0)$
peak. At large values of  $\hat{s}^{1/2}_{min}(0)$, where
a new physics signal is most likely to appear, any SM background 
will be rather smooth and featureless, so that it can be safely
subtracted away through a side-band method.

Another possible application of $\hat{s}^{1/2}_{min}(0)$
is at the trigger level. In Section \ref{sec:compare}
we already saw that $\hat{s}^{1/2}_{min}(0)$ is superior 
to both $H_T$ and $\met$ in identifying the scale of the 
hard scattering. At the same time, there exist dedicated
$H_T$ and $\met$ triggers, motivated by the sensitivity 
of those variables to the relevant energy scale.
Given that our variable is doing an even better job in this respect, 
we believe that the implementation of a high-level 
$\hat{s}^{1/2}_{min}(0)$ trigger should be given a
serious consideration.

As we have been emphasizing throughout, a major advantage of 
$\hat{s}^{1/2}_{min}$ is that it does not require any 
explicit event reconstruction and thus it is very model-independent.
We should mention that to some extent, these properties are also 
shared by the $M_{TGen}$ variable proposed in \cite{Lester:2007fq}.
In calculating $M_{TGen}$, one considers all possible partitions 
of the visible particles $X_i$ in the event, thus effectively 
eliminating the model-dependence which stems from assuming a 
particular topology. While $M_{TGen}$ and $\hat{s}^{1/2}_{min}$ 
are similar in this respect, we believe that $\hat{s}^{1/2}_{min}$ 
has three definite advantages --- first, it is much, much easier to construct.
Second, $\hat{s}^{1/2}_{min}$ can be applied to extreme asymmetric topologies 
where the second side of the event yields no visible particles.
A simple example of this sort was the associated gluino-LSP production 
considered in Figs.~\ref{fig:varscg}, \ref{fig:cg4j},
\ref{fig:ds}(c,d), \ref{fig:meas}(b) and \ref{fig:ds_ISR}(c,d).
Finally, the interpretation of $\hat{s}^{1/2}_{min}$ involves
reading off a peak, while $M_{TGen}$ requires reading off an endpoint.
The former is much easier than the latter: for example, 
a peak would still be recognizable in the presence of 
large backgrounds. In contrast, an $M_{TGen}$ endpoint can fade out 
due to a number of reasons, including detector resolution,
combinatorial background, etc. On the other hand,
$M_{TGen}$ (and more generally, the $M_{T2}$ class of variables) 
is better behaved in the presence of ISR. More specifically,
the {\em endpoints} of the $M_{TGen}$ and $M_{T2}$ distributions
in general do shift in the presence of ISR, and their explicit 
dependence on the ``upstream'' transverse momentum 
has to be calculated on a case by case basis \cite{Burns:2008va}.
However, the nice feature of both $M_{TGen}$ and $M_{T2}$
is that when the test mass $\tilde m_\chi$ becomes equal to its
true value $m_\chi$, there is no such shift and the endpoint
remains intact even in the presence of arbitrary ISR.
In contrast, as discussed in Section \ref{sec:ISR}, 
$\hat{s}^{1/2}_{min}$ is always affected by ISR to some extent, 
requiring some sort of correction.

In conclusion, we reiterate that perhaps the most important advantage of
$\hat{s}^{1/2}_{min}$ is that it is readily available from day one.
We are therefore eagerly looking forward to the first $\hat{s}^{1/2}_{min}$ 
plots produced with real LHC data.

\bigskip

\acknowledgments
We are grateful to A.~Barr, R.~Cavanaugh, R.~Field, A.~Korytov, C.~Lester 
and B.~Webber for useful discussions and correspondence.
This work is supported in part by a US Department of Energy 
grant DE-FG02-97ER41029. Fermilab is operated by Fermi Research Alliance, LLC under
Contract No. DE-AC02-07CH11359 with the U.S. Department of Energy.

%%%%%%%%%%%%%%%%%%%%%%%%%%%%%%%%%%%%%%%%%%%%%%%%%%%%%%%%%%%%%%%
\listoftables           % ONLY DRAFT
\listoffigures          % ONLY DRAFT

%%%%%%%%%%%%%%%%%%%%%%%%%%%%%%%%%%%%%%%%%%%%%%%%%%%%%%%%%%%%%%%

\end{document}